\documentclass[prb,
showpacs,
twocolumn,
floats,
10pt,
aps,
citeautoscript,
longbibliography,
superscriptaddress]{revtex4-2}

\usepackage{placeins}
\usepackage{lipsum}
\usepackage{xcolor}
\usepackage[normalem]{ulem}
\usepackage{comment}
\usepackage{graphicx}
\usepackage{wrapfig}
\usepackage{dcolumn}
\usepackage{bm}

\usepackage{blindtext}
\usepackage{graphics}
\usepackage{verbatim}   
\usepackage[ruled,lined]{algorithm2e}
\usepackage{amsfonts}
\usepackage{amsmath}
\usepackage{amssymb}
\usepackage{mathrsfs} 
\usepackage{multirow}
\usepackage{physics}
\usepackage{bbm}  
\usepackage{adjustbox}
\usepackage{appendix}

\usepackage{tabstackengine}
\setstackEOL{\cr}

\newcommand{\Eq}[1]{Eq.~\eqref{#1}}

\newcommand{\Fig}[1]{Fig.~\ref{#1}}
\newcommand{\Figs}[1]{Figs.~\ref{#1}}

\newcommand{\Sec}[1]{Sec.~\ref{#1}}
\newcommand{\App}[1]{Sup.~\ref{#1}}

\newcommand{\mr}[1]{\mathrm{#1}}
\newcommand{\mc}[1]{\mathcal{#1}}

\newcommand{\z}{z}
\newcommand{\bZero}{{\boldsymbol{0}}}
\newcommand{\bQ}{{\boldsymbol{\mathrm{Q}}}}
\newcommand{\bq}{{\boldsymbol{\mathrm{q}}}}
\newcommand{\bX}{{\boldsymbol{\mathrm{X}}}}
\newcommand{\bY}{{\boldsymbol{\mathrm{Y}}}}

\newcommand{\bK}{{\boldsymbol{\mathrm{K}}}}
\newcommand{\bk}{{\boldsymbol{\mathrm{k}}}}
\newcommand{\br}{{\boldsymbol{\mathrm{r}}}}
\newcommand{\bR}{{\boldsymbol{\mathrm{R}}}}
\newcommand{\bD}{{\boldsymbol{\mathrm{D}}}}
\newcommand{\pdag}{{\phantom{\dag}}}

\newcommand{\pFS}{p_{\mathrm{FS}}}
\newcommand{\pStar}{p^{\ast}}
\newcommand{\pLS}{p_{\mathrm{LS}}}
\newcommand{\pSig}{p_{\Sigma}}

\newcommand{\TFL}{T_{\mr{FL}}}

\newcommand{\LMUMunich}{Arnold Sommerfeld Center for Theoretical Physics, Center for NanoScience, and Munich Center for Quantum Science and Technology, Ludwig-Maximilians-Universit\"at M\"unchen, 80333 Munich, Germany}

\newcommand{\RutgersUniversity}{Department of Physics and Astronomy and Center for Condensed Matter Theory, Rutgers University, Piscataway, New Jersey 08854-8019, USA}

\RequirePackage[
  hyperindex,colorlinks,bookmarksnumbered,
  plainpages=true,pdfstartview=FitH]{hyperref}
\hypersetup{linkcolor=blue,urlcolor=blue,citecolor=blue} 
\usepackage[all]{hypcap}

\newcommand{\TNFL}{T_{\mr{NFL}}}

\begin{document}

\preprint{}

\makeatletter
\def\maketitle{
\@author@finish
\title@column\titleblock@produce
\suppressfloats[t]}
\makeatother

\title{Quantum criticality in the two-dimensional Hubbard model}
\author{Mathias Pelz}
\email{mathias.pelz@lmu.de}
\affiliation{\LMUMunich}

\author{Gabriel Kotliar}
\email{kotliar@physics.rutgers.edu}
\affiliation{\RutgersUniversity}

\author{Jan von Delft}
\email{vondelft@lmu.de}
\affiliation{\LMUMunich}

\author{Andreas Gleis}
\email{andreas.gleis@rutgers.edu}
\affiliation{\RutgersUniversity}

\date{\today}

\begin{abstract}
We study the normal-state, doping-driven phase diagram of the square-lattice Hubbard model using the dynamical cluster approximation combined with the numerical renormalization group as a cluster solver, which gives direct access to real-frequency dynamics at essentially zero temperature.
In a parameter regime relevant for cuprates, $U=7t$ and $t'=-0.3t$, we find a critical doping $\pStar$ that marks a continuous quantum phase transition between a pseudogap metal and a normal Fermi liquid.
The transition is identified by a continuous collapse, from both sides, of the Fermi-liquid scale extracted from charge, spin, and $d_{x^2-y^2}$-wave pairing susceptibilities.
This collapse produces a non-Fermi-liquid regime at intermediate energy scales, which appears to extend to arbitrarily low scales at $\pStar$.
As $\pStar$ is crossed from the normal Fermi liquid at $p>\pStar$ into the pseudogap metal at $p<\pStar$, the coherent low-energy spectral weight in the antinodal region is lost and replaced by a narrow, metallic pseudogap, while the nodal region evolves smoothly and remains comparatively coherent.
This gives rise to Fermi arcs in the pseudogap metal at $p<\pStar$, since the zero-frequency spectral weight remains large in the nodal region but is strongly suppressed in the antinodal region.

\medskip
\noindent
DOI:

\end{abstract}

\maketitle

Due to its relevance to high-$T_c$ cuprates, the problem of doping a Mott insulator~\cite{Lee2006} has been a topic of intense discussion in the past three decades. 
A major mystery is the normal state that emerges upon hole doping in cuprates~\cite{Keimer2015}.
On the overdoped side, transport, thermodynamics, and spectroscopy are broadly consistent with a correlated normal Fermi liquid~(FL). 
By contrast, on the underdoped side, a pseudogap~(PG) metal emerges, featuring a loss of electronic spectral weight predominantly in the antinodal region and a reconstructed Fermi surface incompatible with Luttinger's sum rule.
A central question, raised very early on~\cite{Castellani1995,Varma1997,Varma1999,Tallon1999}, is whether the PG metal and the overdoped FL are separated by a quantum critical point~(QCP) at some critical doping $\pStar$\cite{Timusk1999_Pseudogap,Norman1998,Damascelli2003,He2011,Vishik2012,Reber2019,Sachdev2010}.

Evidence for quantum criticality at the pseudogap endpoint at $\pStar$ has accumulated in several cuprate families. The Hall number, angle-dependent magnetoresistance, and quantum oscillations indicate a transformation of the Fermi surface across $\pStar$, while thermodynamic measurements find a strong enhancement of the specific-heat coefficient near the same regime~\cite{Badoux2016,Collignon2017_PRB,Fang2022_ADMR,DoironLeyraud2007,Sebastian2012,Michon2019}.
Further, the strange-metal region of the phase diagram features several aspects associated with quantum criticality, 
such as $\omega/T$ scaling and anomalous metallic behavior inconsistent with coherent quasiparticles, both within a fan-like shape that appears to emanate from $\pStar$.
Some experimental data also suggests an extended quantum critical region~\cite{Reber2019,Ayres2021}, which has been argued to be a result of disorder~\cite{Patel2024}.

The description of such a PG-QCP poses a major challenge. This is because on the one hand, the PG metal itself is still mysterious and not completely understood, while on the other hand, this QCP, if present, does not seem to be a conventional order-parameter driven QCP describable by Hertz--Millis--Moriya theory~\cite{Hertz1976,Millis1993,Moriya1985}.
Several routes to describe the PG metal and the PG-QCP have been explored, such as the idea of a fractionalized Fermi liquid with topological order that is separated from the normal FL by a deconfined QCP~\cite{Senthil2003,Senthil2004,Moon2011,Scheurer2018,Wu2018,Zhang2020,Mascot2022,Wang2022,Wu2024,Bonetti2026}; fluctuating conventional order~\cite{Emery1995,Sachdev2004,Hayward2014,Nie2015,Eberlein2016,Verret2017,Bonetti2020,Bonetti2022,Klett2022,Li2022,Lihm2026,Forni2026}; or non-local extensions of dynamical mean-field theory~(DMFT) \cite{Georges1996_DMFT,Hettler1998_DCA,Hettler2000_DCA,Maier2005_clusters}. 

In particular, cluster DMFT approaches to the Hubbard model have been very successful in capturing PG metal physics 
\cite{Huscroft2001,Civelli2005_FermiSurfaceBreakup,Maier2005_clusters,Tremblay2006_num,Kyung2006,Stanescu2006_FermiArcs,Macridin2006,Haule2007,Ferrero2009_FermiArcs,Werner2009,Liebsch2009,Mikelsons2009,Vidhyadhiraja2009,Gull2010_patching-pg,Khatami2010,Yang2011,Sordi2010,Sordi2011,Sordi2012_Widom,Sordi2012_PRL,Sordi2013,Sordi2019,Scheurer2018,Wu2018,Meixner2024}.
For instance, these have successfully captured the depression of spectral weight at the Fermi level \cite{Hettler2000_DCA}, the reconstructed Fermi surface due to the presence of Luttinger surfaces \cite{Stanescu2006_FermiArcs}, and the momentum-dependence of spectral weight along the Fermi surface that leads to Fermi arcs and to the nodal--antinodal dichotomy \cite{Civelli2005_FermiSurfaceBreakup}.

Interestingly, it seems that all of this physics can be captured with relatively small four-site clusters, and the qualitative results do not strongly depend on the type of cluster scheme or periodization used.
However, the precise nature of the PG metal, in particular the nature and structure of its low-energy excitations, has not been established so far. 
Furthermore, the existence of a QCP separating the PG metal from the normal FL within cluster DMFT is controversial.
While some work found apparent evidence for a QCP, Sordi \textit{et al.} \cite{Sordi2012_Widom} established that within a certain parameter region ($t'/t=0$, $U/t\leq6$) criticality is associated with the finite-temperature endpoint of a first-order transition.

The main limitation in establishing the existence or not of a PG--QCP in a relevant parameter regime of the Hubbard model within cluster DMFT has been insufficient low-energy and low-temperature resolution, and the inability to access the relevant parameters directly.
In this work, we overcome this limitation by using four-patch DCA together with the numerical renormalization group~(NRG) as a cluster impurity solver.
NRG can resolve exponentially low energy and temperature scales and can access arbitrary parameters, though NRG is limited by Hilbert space size.
We therefore deliberately restrict calculations to the normal state and do not consider symmetry-broken solutions. For details on the methods see Sec.~\ref{sec:methods} and Supplement Material \ref{app:model&method}.

To describe cuprate physics, we use the square lattice Hubbard model with Hamiltonian
\begin{align}
    \nonumber H &=  \sum_{\br\br'\sigma}  -t_{\br\br'}\,c_{\br\sigma}^\dag c^{\pdag}_{\br'\sigma} -  \mu\sum_{\br\sigma} n_{\br\sigma} +  U\sum_{\br} n_{\br\uparrow} n_{\br\downarrow}\\
    &= \sum_{\bk\sigma} (\epsilon_{\bk}-\mu)\,n_{\bk\sigma}  +  U\sum_{\br} n_{\br\uparrow} n_{\br\downarrow}\ ,
\end{align}
where $c^{\pdag}_{\br\sigma}$ annihilates a spin-$\sigma$ electron at site $\br$, $n_{\br\sigma} = c^{\dag}_{\br\sigma} c^{\pdag}_{\br\sigma}$.
For the hopping matrix elements $t_{\br\br'}$, we consider nearest and next-nearest neighbor terms, denoted $t$ and $t'$. We set $t = 1$ as our unit of energy and choose $t'/t = -0.3$ and $U/t = 7$, suitable for describing cuprate physics. We tune $\mu$ to adjust the hole-doping $p$ (filling $n = 1-p$). We study this model using four-patch DCA+NRG (see \Sec{app:model&method} of the supplemental material~(SM)~\cite{supplement}) and analyze its physics by computing several real-frequency
correlators involving local or short-ranged operators defined on the cluster, see Sec.~\ref{sec:methods} on Methods.

\begin{figure}[t!]
    \centering
    \includegraphics[width=1\linewidth]{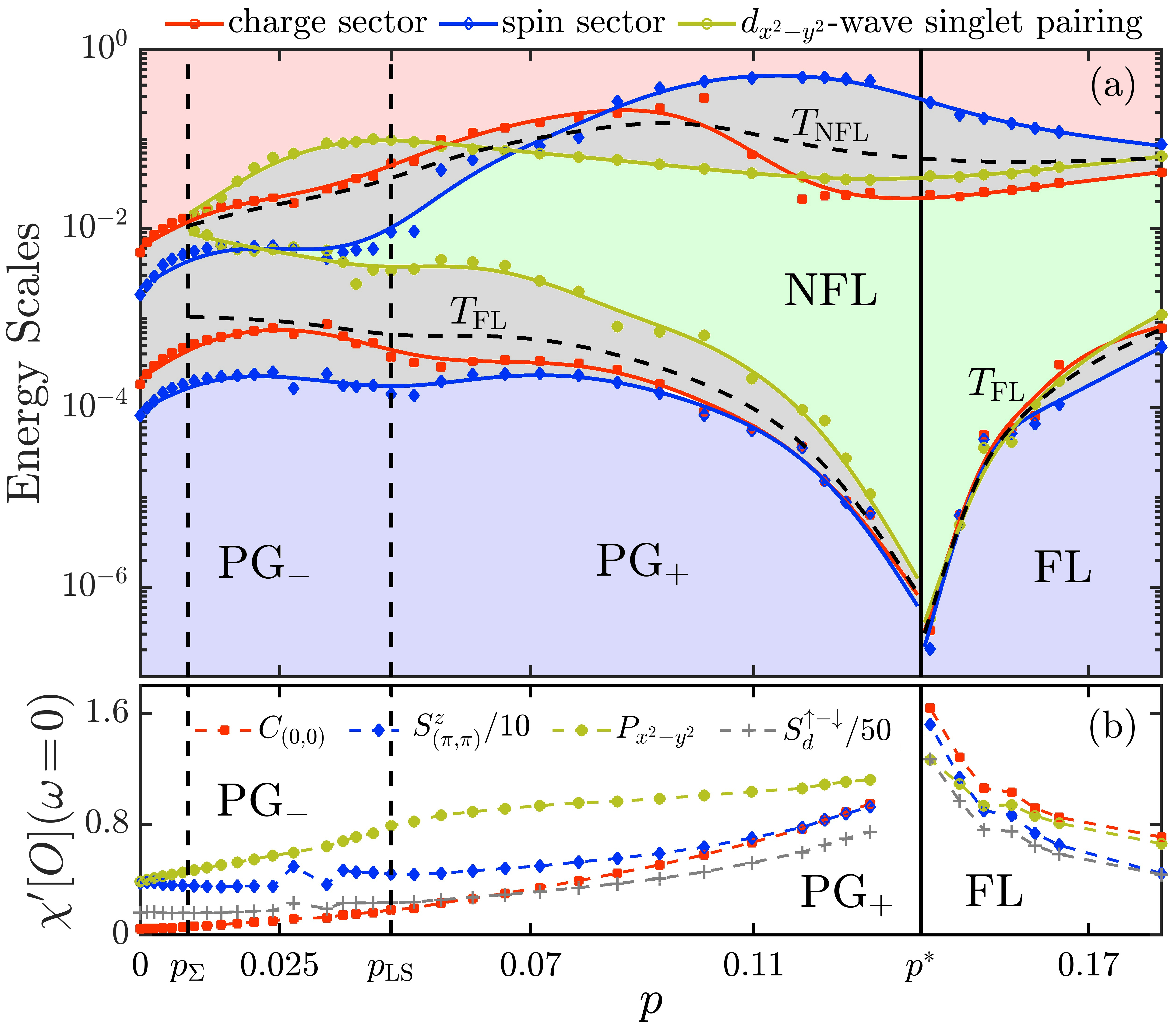}
    \caption{Doping-driven, zero-temperature normal-state phase diagram and static susceptibilities of the square lattice Hubbard model with $U = 7t$ and $t'=-0.3t$, obtained with DCA+NRG. 
    (a) Crossover energy scales (dots) extracted from the frequency dependence of dynamical cluster susceptibilities $\chi[O](\omega)$ for operators $O$ describing charge (red), spin (blue) and $d$-wave pairing (green) [see Eqs.~\eqref{eq:operators_main}].  Colored lines are guides-to-the-eye.
    The geometric averages of the lower-lying or upper-lying three crossover scales are indicated  by black curves marked $T_{\mr{FL}}$ or $T_{\mr{NFL}}$, respectively. For $p > 0$, we find FL behavior for energies below $\TFL$, and non-Fermi-liqui (NFL) behavior for energies between 
    $\TFL$ and $\TNFL$. There are several low-energy FL phases: a normal FL for $p > \pStar$ and a PG metal for $p < \pStar$, separated by a QCP at $\pStar=0.14$ (vertical solid line) where $T_{\mr{FL}}$ is suppressed to zero. The PG metal comprises two sub-regimes,  PG$_-$ and PG$_+$, which are not separated by a phase transition (see the text for definitions of $\pLS$ and $\pSig$, vertical dashed lines).
    (b) Selected static susceptibilities $-\chi'[O](\omega= 0)$ as functions of doping $p$.}
    \label{fig:t'-0.3_phasedia}
\end{figure}

\section{Results}
\subsection{Phase diagram}

\begin{figure}[t!]
    \centering
    \includegraphics[width=1\linewidth]{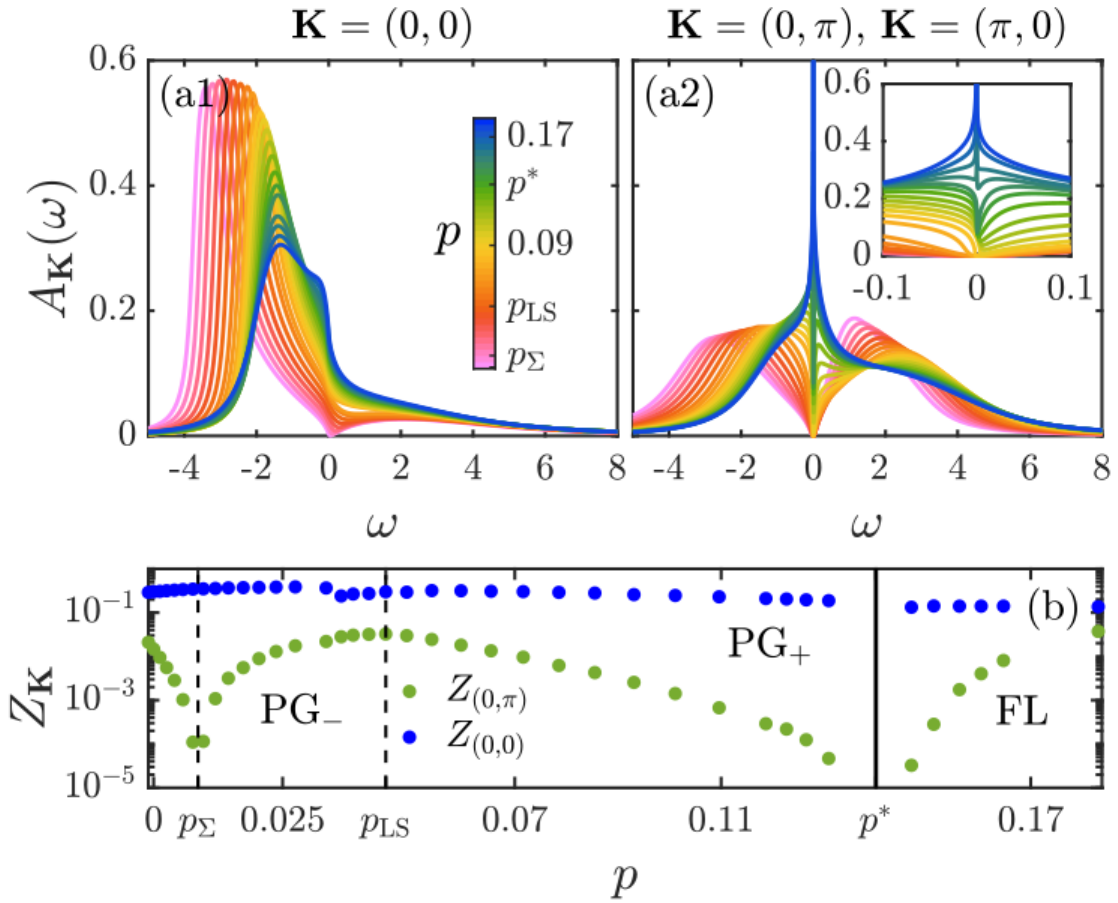}
    \caption{Patch spectral functions and quasiparticle weights. (a1,a2) Cluster spectral functions $A_{\bK}(\omega)$ of the nodal patch $\bK=(0,0)$ and the antinodal patches $\bK=(0,\pi)$, equivalently $\bK=(\pi,0)$. The inset shows details of the low-frequency region around $\omega=0$; colors indicate doping. Across $\pStar$, the antinodal spectrum changes from a resonance for $p>\pStar$ to a pseudogap for $p<\pStar$, whereas the nodal spectrum evolves smoothly. (b) Corresponding quasiparticle weights $Z_{\bK}=[1-\partial_\omega\mr{Re}\,\Sigma_{\bK}(\omega)|_{\omega=0}]^{-1}$ versus doping for the same patches. Vertical lines mark $\pStar$, $\pLS$, and $\pSig$. The antinodal $Z_{(0,\pi)}$ is suppressed to zero at $\pStar$ and again at $\pSig$; the latter occurs because the antinodal self-energy pole crosses $\omega=0$. The nodal $Z_{(0,0)}$ remains finite within numerical resolution. The $\bK=(\pi,\pi)$ patch is omitted because it remains gapped.}
    \label{fig:cq_dual}
\end{figure}

Figure~\ref{fig:t'-0.3_phasedia}(a) shows the doping-driven, zero-temperature normal-state phase diagram obtained from DCA+NRG.
Since symmetry breaking is excluded, all regimes shown are normal-state solutions.
For $p>0$, we find two metallic phases: a normal FL for $p>\pStar$ and a PG metal for $p<\pStar$.
The critical doping $\pStar=0.14$ is the only point in this part of the phase diagram where we find a genuine quantum phase transition.
Its central signatures are (i) the continuous suppression of the FL coherence scale as $\pStar$ is approached from either side, and (ii) the abrupt change of the low-frequency cluster spectral functions in the antinodal patches, $\bK=(0,\pi)$ and $\bK=(\pi,0)$, from exhibiting a narrow metallic pseudogap for $p<\pStar$ to a sharp resonance for $p>\pStar$  [cf.~Fig.~\ref{fig:cq_dual}, discussed later].

The markers $\pLS$ and $\pSig$ at lower doping in \Fig{fig:t'-0.3_phasedia} have a different status: rather than additional quantum phase transitions, they denote crossovers within the PG metal, 
associated with changes in the Luttinger-surface and self-energy-pole structure familiar from cluster studies of the pseudogap~\cite{Stanescu2006_FermiArcs,Ferrero2009_FermiArcs,Gull2010_patching-pg,Haule2007}.
Upon increasing $p$ from the Mott-insulating side, the antinodal gap starts to fill when $p$ rises above $\pLS$; equivalently, upon decreasing $p$ from within the $\mr{PG}_+$ regime, 
the partially filled antinodal pseudogap becomes fully gapped
when $p$ drops below $\pLS$.
The doping $\pSig$ denotes the point where the antinodal self-energy pole crosses the Fermi level.
Neither $\pLS$ nor $\pSig$ is accompanied by a suppression of the FL scale.
In the discussion below, we therefore first focus on the $\mr{PG}_+$ and normal FL regimes ($p > \pLS$), in particular the vicinity of $\pStar$, before briefly returning to these two crossovers at lower doping.

\begin{figure}[t!]
    \centering
    \includegraphics[width=1\linewidth]{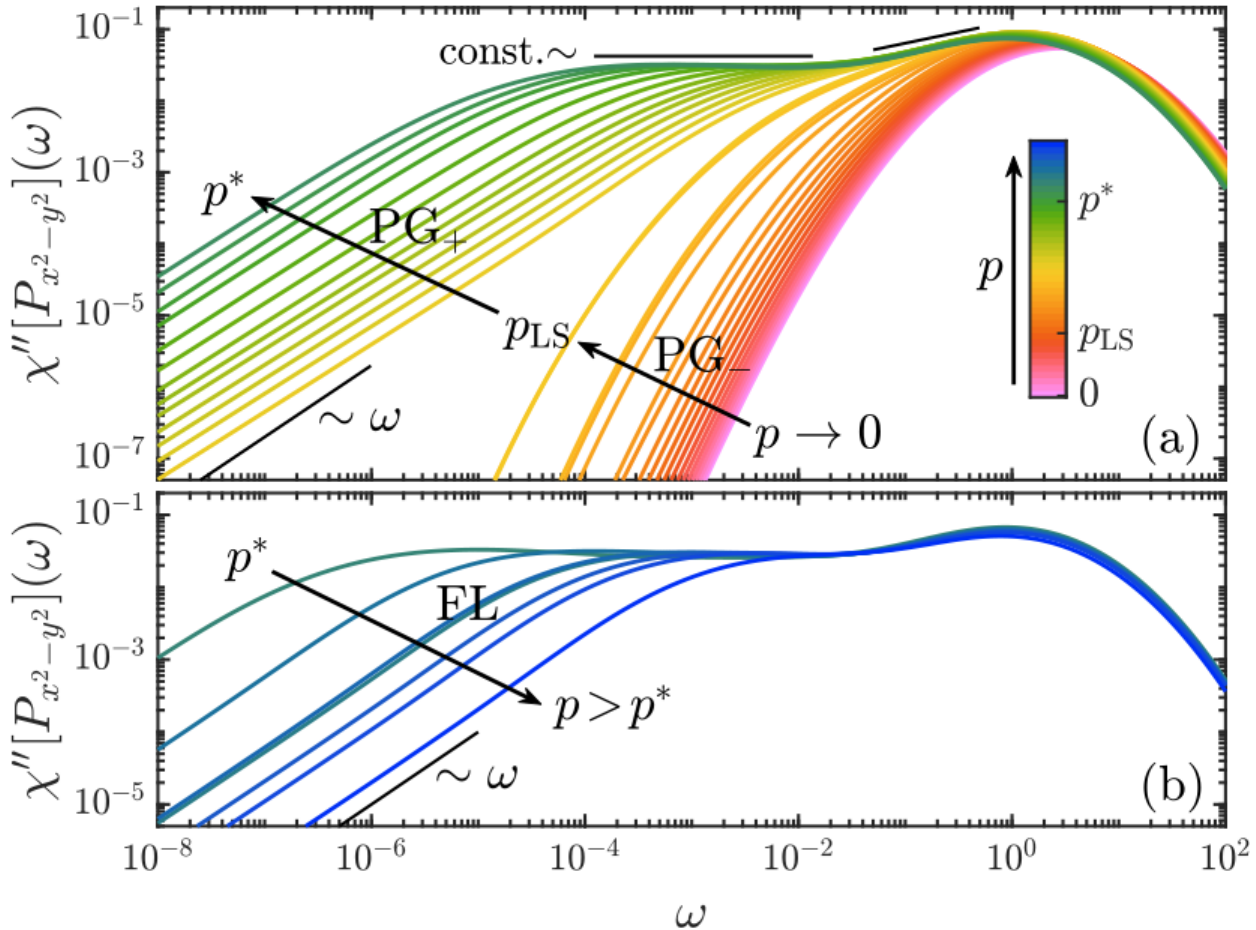}
    \caption{Doping evolution of the local $d_{x^2-y^2}$-wave singlet-pair spectrum $\chi''[P_{x^2-y^2}](\omega)$ on a log--log scale. (a) Evolution from small doping in the PG$_-$ regime (pink), through $\pLS$, and toward $\pStar$ (green). (b) Evolution from $\pStar$ into the overdoped FL regime (blue). Arrows indicate increasing $p$. The guide-to-the-eye lines mark the low-frequency FL behavior, $\chi''\propto\omega$, and the intermediate-frequency NFL plateau $\chi''\propto\mr{const.}$; the crossover between them defines the FL scale $\TFL$. Approaching $\pStar$ from either side, this scale is suppressed, and the plateau extends to ever lower frequencies. The upper end of the plateau defines the NFL scale $\TNFL$.}
    \label{fig:chiP_doping}
\end{figure}

To identify the quantum phase transition at $\pStar$, we track energy scales extracted from dynamical susceptibilities of local charge, spin and $d$-wave pairing operators.
The corresponding operators and computational details are discussed in Sec.~\ref{sec:methods}. As a representative example, Fig.~\ref{fig:chiP_doping} shows the doping dependence of $\chi''[P_{x^2-y^2}](\omega)$, the spectral part of the local $d$-wave pair susceptibility. At frequencies below a crossover scale $\TFL$,  $\chi''(\omega)$ shows linear-in-$\omega$ behavior typical of a FL. We have checked that the NRG finite-size many-body  are indeed Fermi liquid spectra for both $p > \pStar$ and $p < \pStar$. The FL scale is continuously suppressed to zero at $\pStar$, signaling a QCP governed by a non-Fermi liquid fixed point.
Figure~\ref{fig:t'-0.3_phasedia}(b) shows $\chi'(\omega=0)$, the real part of the static charge, spin and pairing susceptibilities; all exhibit a pronounced peak and possibly a divergence at $\pStar$.

Returning to Fig.~\ref{fig:chiP_doping} for $\chi''[P_{x^2-y^2}](\omega)$ as a representative example, 
we next describe how we identified the various crossover energy scales and regimes. Starting from high frequencies, $\chi''$ exhibits a broad maximum around $\omega\simeq 1$ and decreases toward lower frequencies.
Near $\pStar$, an intermediate-frequency plateau emerges, $\chi''(\omega)\propto \mr{const.}$, reminiscent of bosonic spectra in the non-Fermi-liquid regimes of, for instance, the two-channel and two-impurity Kondo models.
At even lower frequencies, this plateau crosses over to FL behavior, with $\chi''(\omega)\propto\omega$.
As $\pStar$ is approached from either side, the FL crossover scale is pushed to ever lower frequencies and the plateau correspondingly extends downward. At the QCP, our data suggest that the plateau persists to $\omega\to0$, which would imply a logarithmical divergence for $\chi'(\omega=0)$, the real part of the static susceptibility.
The evolution with doping of $\chi''[P_{x^2-y^2}](\omega)$ across $\pStar$ is otherwise smooth, consistent with a continuous quantum phase transition.

\subsection{Single-particle spectra}

The same transition is visible in the single-particle spectra.
Figures~\ref{fig:cq_dual}(a1) and \ref{fig:cq_dual}(a2) show the cluster spectral functions $A_{\bK}(\omega)$ of the patches $\bK = (0,0)$ and $\bK = (0,\pi)$ [or equivalently $\bK = (\pi,0)$], respectively. These correspond to the momentum-averages of the lattice spectral functions in the nodal and antinodal regions, respectively.
For $p > \pLS$, the Fermi surface is located within in these patches, and the corresponding patch spectral functions are therefore metallic, i.e.\ $A_{\bK}(\omega = 0) > 0$. 
The $\bK = (\pi,\pi)$ patch is not shown, since it remains gapped for all dopings considered.

While $A_{(0,0)}$ [Fig.~\ref{fig:cq_dual}(a1)] evolves smoothly with doping through the transition, $A_{(0,\pi)}$ [Fig.~\ref{fig:cq_dual}(a2)] changes sharply at the Fermi level.
For $p>\pStar$, in the normal FL, $A_{(0,\pi)}$ contains a narrow resonance at $\omega=0$ [inset of Fig.~\ref{fig:cq_dual}(a2)], which sharpens as $p$ is lowered toward $\pStar$.
For $p<\pStar$, this resonance gives way to narrow PG.
This PG arises due to a pole in the antinodal self-energy that emerges at $p < \pStar$, see Fig.~\ref{fig:SpecFunc}(b2) of the SM~\cite{supplement}.
In the $\mr{PG}_+$ regime ($\pLS \!<\! p \!<\! \pStar$), the PG is still partially filled, $A_{(0,\pi)}(\omega = 0)>0$, so the antinodal patches still contribute quasiparticle excitations and a Fermi-surface segment.

\begin{figure*}[t!]
    \centering
    \includegraphics[width=1\linewidth]{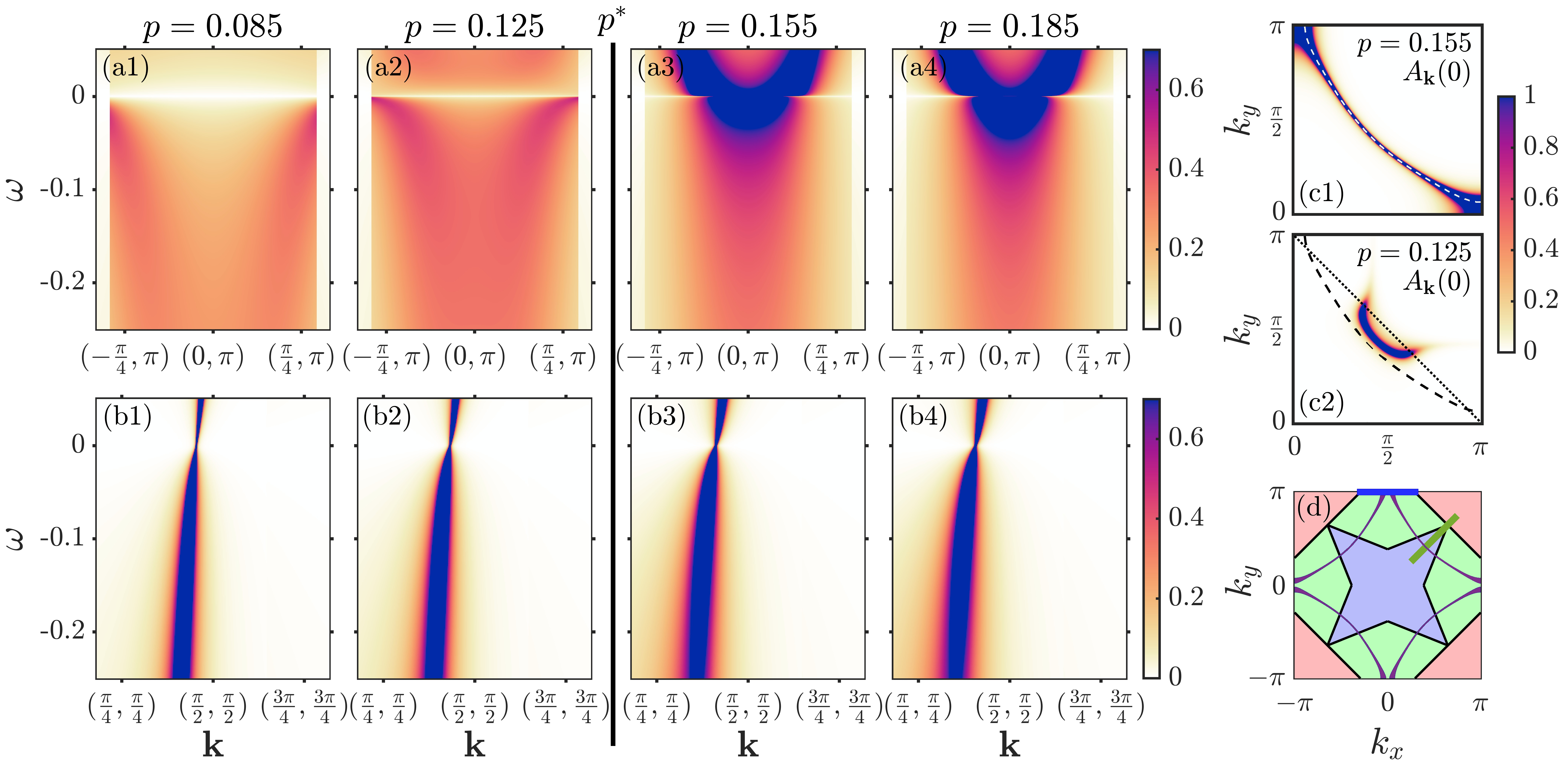}
    \caption{Momentum-resolved spectral functions on both sides of the pseudogap QCP at $\pStar$.
    (a,b) Non-interpolated DCA spectra $A_{\bk}(\omega)$ along antinodal and nodal cuts [indicated in (d)], respectively. 
    Columns show dopings below and above $\pStar\simeq0.14$; the black vertical separator marks $\pStar$.
    For the antinodal cut (a), the low-energy spectral weight present on the FL side ($p > \pStar$) is strongly depleted on the PG side ($p < \pStar$), whereas for the nodal cut (b) the low-energy spectrum evolves smoothly.  (c1,c2) $\mc{L}$-interpolated zero-frequency spectral function $A_{\bk}(0)$ in the FL at $p=0.155>\pStar$ and in the PG$_+$ regime at $p=0.125<\pStar$, respectively.
    In (c1), the curved dashed line denotes the locus of maximal $A_{\bk}(0)$; it is shown in (c2), too, as a reference for the unreconstructed Fermi surface; the straight dashed line in (c2) marks the antiferromagnetic zone boundary.
    (d) DCA patch geometry, showing the antinodal (thik blue) and antinodal (thick green) momentum cuts used for (a) and (b), respectively, and the Fermi surface (purple) depicted in (c1).
    }
    \label{fig:ARPES_A0}
\end{figure*}

Figure~\ref{fig:cq_dual}(b) shows the corresponding quasiparticle weights,
$Z_{\bK}=[1-\partial_\omega \mr{Re}\,\Sigma_{\bK}(\omega)|_{\omega=0}]^{-1}$,
obtained from a numerical derivative of the corresponding real part of the patch self-energy.
On approaching the QCP from either side, $Z_{(0,\pi)}$ is suppressed to zero.
This is consistent both with the change of the low-frequency structure of $A_{(0,\pi)}(\omega)$ and with the collapse of the FL scale in the cluster susceptibilities.

By contrast, $Z_{(0,0)}$ appears to remain finite across the transition.
Taken at face value, this would suggest a momentum-selective 
NFL at the QCP.
However, this apparent patch selectivity should not be overinterpreted: it is difficult to reconcile a non-zero nodal quasiparticle weight with the vanishing FL scale that we observe for several cluster susceptibilities at the QCP [cf.~\Fig{fig:t'-0.3_phasedia}(a)].
Following arguments from marginal FL~(MFL) phenomenology~\cite{Varma1989_MFL}, scattering of the $\bK=(0,0)$ electrons off short-ranged critical bosonic fluctuations as found, for instance, in Fig.~\ref{fig:chiP_doping} 
should generate at least a logarithmically weak suppression
of $Z_{(0,0)}$, 
with $Z_{(0,0)}^{-1}$ growing as $\ln(1/\TFL)$ close to the QCP.
However, resolving such a logarithmic dependence requires highly accurate numerics extremely close to $\pStar$.
In a follow-up paper~\cite{Pelz2026_finiteT}, where we examine quantum critical scaling in the finite-temperature NFL region, we find evidence for MFL physics in the $\bK=(0,0)$ patch from an $\omega/T$ scaling analysis.
A definitive zero-temperature statement on the nodal quasiparticle weight will require additional methodological improvements and is left for future work.

The PG metal below $\pStar$ is further subdivided into $\mr{PG}_+$ for $\pStar>p>\pLS$ and $\mr{PG}_-$ for $p<\pLS$.
This subdivision reflects changes in the antinodal spectral weight that occur smoothly, not an additional quantum phase transition.
In $\mr{PG}_+$ the antinodal pseudogap is partially filled, whereas at $\pLS$ it becomes fully gapped upon lowering $p$, see Fig.~\ref{fig:cq_dual}(a2) and its inset.
Equivalently, upon increasing the doping from the Mott insulator at $p = 0$, $\pLS$ is the point where antinodal spectral weight at the Fermi level first appears.
Thus, in $\mr{PG}_-$ only the $\bK=(0,0)$ patch remains metallic; the state is a patch-selective metal with insulating antinodal patches.
Because the full gap in $A_{(0,\pi)}(\omega)$ opens smoothly and the FL scale is not suppressed at $\pLS$, this crossover is distinct from the QCP at $\pStar$.
The mechanism for the full antinodal gap is analogous to that of Ref.~\cite{Ferrero2009_FermiArcs} and is discussed further in Fig.~\ref{fig:ReSE0} of the SM~\cite{supplement}.

The full antinodal gap for $p<\pLS$ leads to a suppression of the low-frequency spectral weight of bosonic dynamical susceptibilities.
In particular, the local $d$-wave pair spectrum decreases faster than linear-in-$\omega$ at low frequencies [\Fig{fig:chiP_doping}(a)], reflecting the absence of antinodal quasiparticle excitations.
Other susceptibilities, such as the uniform charge susceptibility on the cluster (see Fig.~\ref{fig:chiC_doping} of the SM~\cite{supplement}), remain linear in $\omega$, but with a much smaller slope than for $p>\pLS$.

As the doping is reduced further, a more subtle crossover occurs at $\pSig$.
There, the antinodal self-energy pole that suppresses $A_{(0,\pi)}(\omega)$ near $\omega=0$ crosses from positive to negative frequency.
This crossing produces a zero of $Z_{(0,\pi)}$
[\Fig{fig:cq_dual}(b)], because $\Sigma_{(0,\pi)}(0)$ diverges at $\pSig$. It does not produce a new feature in the spectral function, though, since $A_{(0,\pi)}(\omega)$ is already fully gapped at this point.
Thus,  in stark contrast to the collapse $Z_{(0,\pi)} \to 0$ at $\pStar$, the zero of $Z_{(0,\pi)}$ at $\pSig$ does not signal the destruction of existing QPs. 

For completeness, we have mapped out the phase diagram for several values of the next-nearest-neighbor hopping, $t'/t\leq 0$, at fixed $U=7t$; the resulting characteristic dopings are summarized in Fig.~\ref{fig:t'_doping} of the SM~\cite{supplement}.
The dopings $\pStar$, $\pLS$, and $\pSig$ all shift quantitatively with $t'$, but the overall physical picture remains qualitatively unchanged. This indicates that the discussion above is not special to the representative choice $t'=-0.3t$, but is a robust feature of the square-lattice Hubbard model within the parameter range studied.

\subsection{Fermi surface reconstruction}

As discussed above, the QCP is marked by a sharp change in the  single-particle spectra of the antinodal patches: the $\bK=(0,\pi)$ and $\bK=(\pi,0)$ spectra lose their low-energy FL resonance and develop a narrow pseudogap, while the $\bK=(0,0)$ spectrum evolves smoothly across $\pStar$. This is the cluster analogue of the ARPES phenomenology of the cuprate PG, where low-energy spectral weight is strongly depleted near the antinode when entering the PG regime, while Fermi arcs survive around the node~\cite{Norman1998,Damascelli2003,He2011,Vishik2012,Reber2019,Chen2019}.

These signatures of cluster spectra have a direct momentum-resolved manifestation. Figure~\ref{fig:ARPES_A0} shows how
the momentum-resolved spectral functions, $A_{\bk}(\omega)$, evolve with doping across the QCP. Rows (a) and (b) show non-interpolated DCA spectra along antinodal and nodal cuts [indicated in panel (d)], respectively. 
Along the antinodal cut, the coherent low-energy weight present for $p>\pStar$ is strongly depleted once $p<\pStar$, consistent with the pseudogap in the antinodal patch spectral function of Fig.~\ref{fig:cq_dual}(a2) and with the suppression of $Z_{(0,\pi)}$ in Fig.~\ref{fig:cq_dual}(b).
Along the nodal cut, by contrast, the dispersive low-energy feature 
changes only weakly with doping, in line with the smooth evolution of $A_{(0,0)}(\omega)$.
A phenomenologically similar doping-driven evolution of spectral weight is found experimentally in Bi2212~\cite{Reber2019,Chen2019}.

The spectral functions at the Fermi level at $p = 0.125 < \pStar$ ($\mr{PG}_+$) and $p = 0.155 > \pStar$ (FL) are shown in Figs.~\ref{fig:ARPES_A0}(c1,c2).
They are computed using an interpolated self-energy
$\Sigma_{\bk}(\omega)$ obtained using Liouvillian
($\mc{L}$) interpolation~\cite{Pelz2026_Linterp}.
At $p = 0.155 > \pStar$ in the normal FL, the spectral weight is distributed evenly along the Fermi surface [dashed line in Fig.~\ref{fig:ARPES_A0}(c1)].
Further, the Fermi surface volume is consistent with the Luttinger sum rule. Its shape roughly resembles that of the free system, since the self-energy is only weakly momentum dependent.
By contrast, at $p = 0.125 < \pStar$ in the $\mr{PG}_+$ region, the spectral weight depends strongly on the position along the Fermi surface: it exhibits prominent spectral weight in the nodal region and vanishingly small spectral weight in the antinodal region, leading to Fermi arcs which seem to terminate at the antiferromagnetic zone boundary.

\begin{figure}
    \centering
    \includegraphics[width=1\linewidth]{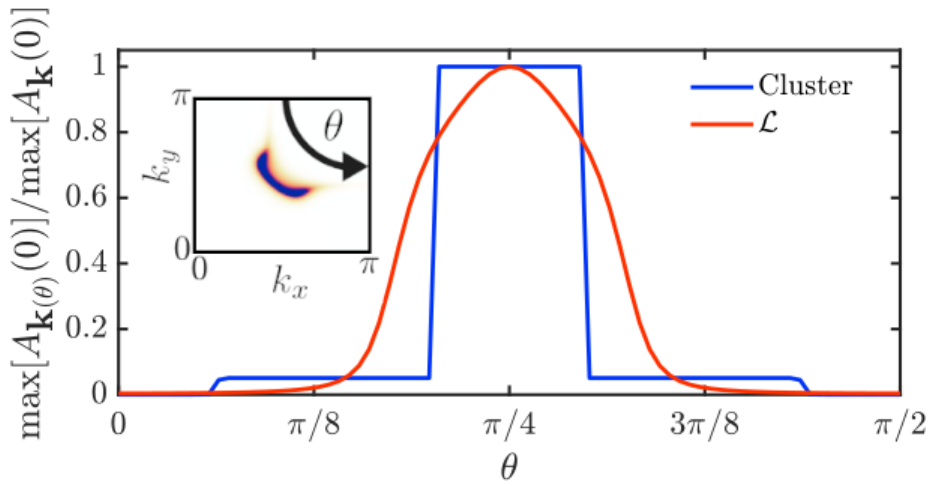}
    \caption{Angular distribution of the zero-frequency spectral weight along the Fermi surface in the PG$_+$ regime at $p=0.125$.
    For each ray $\ell_\theta$ emanating from $(\pi,\pi)$
    at a specified angle $\theta$ as indicated in the inset, we plot $\max_{\bk\in\ell_\theta} A_{\bk}(\omega = 0)$, normalized by its maximum value.
    Blue: the direct DCA cluster result, with abrupt steps at the patch boundaries; red: Liouvillian ($\mc{L}$) interpolation smooths out the steps. 
    }
    \label{fig:maxAk_theta}
\end{figure}

Figure~\ref{fig:maxAk_theta} quantifies this 
by showing the distribution of the zero-frequency spectral weight along the Fermi surface in the PG$_+$ regime. This quantity corresponds to that measured in ARPES studies aimed at analyzing the quasiparticle weight (see, e.g.~Refs.~\cite{Kanigel2006,Reber2012_ARPES_cuprates}).
The cluster result (blue) exhibits a step-like profile, while 
Liouvillian interpolation (red) provides a smooth interpolation.
Similar to experimental findings~\cite{Kanigel2006,Reber2012_ARPES_cuprates}, we find a maximum around $\theta = \pi/4$ (nodal region), with strongly decaying flanks approaching zero around $|\theta-\pi/4| \gtrsim \pi/8$ (antinodal region).

\section{Conclusion}

Using DCA+NRG, we have mapped out the doping-driven, zero-temperature normal-state phase diagram of the square-lattice Hubbard model at $U=7t$, focusing on $t'=-0.3t$ (for other values of $t'/t\leq0$, see the SM~\cite{supplement}).
For hole doping $p>0$, we find two distinct metallic phases: a PG metal for $p<\pStar$ and a normal FL for $p>\pStar$.
These are separated by a QCP at the critical doping $\pStar$, where the FL scale $\TFL$ is suppressed to zero. At intermediate energy scales in the vicinity of the QCP, various dynamical susceptibilities display NFL behavior. We thus settle a long-standing open question: the NFL behavior observed in DMFT treatments of this model
indeed does stem from a QCP. We provide a detailed finite-temperature analysis of the NFL regime in Ref.~\cite{Pelz2026_finiteT}.

In future work, it would be interesting to make a more detailed connection with previous cluster DMFT studies of the doped Hubbard model~\cite{Civelli2005_FermiSurfaceBreakup,Stanescu2006_FermiArcs,Haule2007,Sordi2012_Widom,Ferrero2009_FermiArcs,Gull2010_patching-pg}, in particular with the work of Sordi \textit{et al.}~\cite{Sordi2012_Widom}.
That work showed that for $t'=0$ and $U/t\leq6$, the PG-to-normal FL transition is first-order, and criticality is associated with a finite-temperature critical endpoint.
In the parameter regime studied here, we do not find evidence for a first-order transition; instead, the FL scale appears to vanish continuously at $\pStar$ and physical observables evolve smoothly across $\pStar$. 
Nevertheless, we cannot exclude a very narrow coexistence region, or a critical endpoint at exponentially low temperatures.
It would therefore be useful to use cluster DMFT+NRG to bridge the gap between the regime studied here and that of Sordi \textit{et al.}~\cite{Sordi2012_Widom}.

\section{Methods} \label{sec:methods}

To obtain approximate normal-state solutions of the doped square-lattice Hubbard model, we employ a four-patch DCA scheme,
which approximates the momentum dependence of the self-energy as piecewise constant within patches.
These are labeled by the momenta at the center of the patch, $\bK_1=(0,0)$, $\bK_2=(0,\pi)$, $\bK_3=(\pi,0)$ and $\bK_4=(\pi,\pi)$. We use the ``star patching geometry'' of Ref.~\cite{Gull2010_patching-pg} (see \Fig{fig:ARPES_A0}(d), and Sec.~\ref{app:model&method} of the SM~\cite{supplement} for details), 
which includes the nodal points $\bk=(\pm\pi/2,\pm\pi/2)$
in the patch centered around $\bK_1$. For details on the numerical evaluations of local lattice Green's functions and hybridization functions, see Sec.~\ref{app:AsyEst_LatInt} of the SM~\cite{supplement}.

The cluster impurity model is solved using the MuNRG implementation \cite{Lee2016_NRG,Lee2017} of NRG \cite{Wilson1975_NRG,Krishna-murthy1980_NRG,Anders2005_NRG,Bulla2008_NRG,Kugler2022_SEtrick}, based on the QSpace tensor library \cite{Weichselbaum2012a_QSpace,Weichselbaum2012b_QSpace,Weichselbaum2020_QSpace,Weichselbaum2024_QSpace,Weichselbaum2024_QSpaceCode}. We exploit U(1) charge and SU(2) spin symmetry, use a discretization parameter $\Lambda=8$ without $z$-shifting. We retain up to $N_{\mathrm{keep}}=3\cdot 10^4$ U(1)$\times$SU(2) multiplets, which results in truncation energies $E_{\mr{trunc}}\geq5$. We further employ a fully interleaved Wilson chain geometry \cite{Mitchell2014_iNRG,Stadler2016_iNRG} with bath modes ordered as $\bK_1=(0,0)$, $\bK_2=(\pi,0)$, $\bK_3=(0,\pi)$, and $\bK_4=(\pi,\pi)$. Discrete NRG data are broadened using a log-Gaussian scheme \cite{Lee2016_NRG} with broadening parameter $\sigma=0.7$, and self-energies $\Sigma_{\bK}(\omega)$ are computed using Kugler’s symmetric estimator approach \cite{Kugler2022_SEtrick}. Although interleaving formally breaks the symmetry between the $(\pi,0)$ and $(0,\pi)$ patches, the resulting self-energies differ by less than $10^{-3}$. We therefore enforce the symmetry in the DMFT loop by setting $\Sigma_{(0,\pi)}(\omega)\leftarrow\Sigma_{(\pi,0)}(\omega)$ in each iteration.

We analyze the physics of this model both through electronic spectral functions and by computing real-frequency dynamical susceptibilities for several local or short-ranged operators defined on the cluster. 
Susceptibilities are defined by
\begin{align}
    \label{eq:chi}
    \chi[O](\omega) = -\mr{i} \int_0^{\infty} \mr{d}t\, \mr{e}^{\mr{i}(\omega + \mr{i}0^+) t}\langle [O(t),O^{\dagger}(0)]\rangle ,
\end{align}
where $O$ is some bosonic operator acting non-trivially only on the cluster, and $\langle \;\; \rangle$ denotes the thermal expectation value. 
We compose them as $\chi = \chi' - \mr{i} \pi \chi''$
into a real part $\chi'$ and a spectral part $\chi''$.
Within NRG, we compute discrete spectral functions, broaden them with a log-Gaussian, and then compute $\chi'$ via Kramers-Kronig.
In the vicinity of fixed points, spectral functions of dynamical susceptibilities usually exhibit power-law behavior, i.e.\ straight lines on a log-log scale. 
We infer crossover energy scales by inflection points of the spectra on the log-log scale (see \Fig{fig:chi_extraction} of Ref.~\cite{supplement}).

The operators discussed in the main text, whose energy scales are presented in Fig.~\ref{fig:t'-0.3_phasedia}, are defined as:
\begin{subequations}
\label{eq:operators_main}
\begin{align}
\label{eq:Pxy}
        C_{\bQ} &= \sum_{\bK,\sigma \sigma'} c^{\dagger}_{\bK+\bQ,\sigma} \delta^{\pdag}_{\sigma\sigma'} c^{\pdag}_{\bK,\sigma'} \, ,
\\
    S^{z}_{\bQ} & = \sum_{\bK,\sigma\sigma'}  c^{\dagger}_{\bK+\bQ,\sigma} \sigma^{z}_{\sigma\sigma'} c^{\pdag}_{\bK,\sigma'} \, ,
\\ 
P_{x^2-y^2} &= \sum_{\sigma \sigma'} \bigl( c^{\pdag}_{\delta\bX,\sigma} - c^{\pdag}_{\delta\bY,\sigma} \bigr) \mr{i}\sigma^y_{\sigma\sigma'} c^{\pdag}_{\bZero,\sigma'} \, ,
\\
 S^{\uparrow-\downarrow}_{d} &= \sum_{\sigma \sigma'} 
 (c^{\dagger}_{\delta\bD,\sigma} - c^{\dagger}_{\bZero,\sigma})\sigma^{z}_{\sigma\sigma'} (c^{\pdag}_{\delta\bD,\sigma'} - c^{\pdag}_{\bZero,\sigma'}) 
\end{align}
\end{subequations}
Here, $\sigma^z$ and $\sigma^y$ are Pauli matrices in the spin sector, 
$\bK$, $\bQ$ are DCA patch momenta, and $\delta\bX$, $\delta\bY$ $\delta \bD = \delta \bX + \delta \bY$ denote real-space displacements on the cluster relative to its site $\boldsymbol{0}$. For  \Fig{fig:t'-0.3_phasedia}(a), 
the crossover scales in the charge and spin sectors 
were extracted from susceptibilities $\chi''[C_\mr{loc}]$
and $\chi''[S^z_\mr{loc}]$ involving the local charge and spin operators, 
\begin{align}
\label{eq:ClocSloc}
C_\mr{loc} = \tfrac{1}{\sqrt{4}} \sum_{\bQ} C_{\bQ} \, , 
\qquad 
S^z_\mr{loc} = \tfrac{1}{\sqrt{4}} \sum_{\bQ} S^z_{\bQ} \, . 
\end{align}

Further operators for which we computed susceptibilities [\Figs{fig:chi_various}, \ref{fig:chi_extraction}] and details on the extraction of energy scales [\Fig{fig:chiC_doping}] are provided in  Supplement~\ref{app:sus} \cite{supplement}.

\section*{Acknowledgments}
We thank Antoine Georges, Seung-Sup Lee and Alexander Lichtenstein for valuable discussions. We acknowledge the Gauss Centre for Supercomputing e.V. (www.gauss-centre.eu) for funding this project by providing computing time on the GCS Supercomputer SUPERMUC-NG at Leibniz Supercomputing Centre (www.lrz.de). We also acknowledge additional computational resources provided by the Arnold Sommerfeld Center for theoretical physics (www.theorie.physik.uni-muenchen.de). This work was supported in part by the Deutsche Forschungsgemeinschaft under grants INST 86/1885-1 FUGG, LE 3883/2-2, DE 730/16-2, and Germany’s Excellence Strategy EXC-2111 (Project No. 390814868). It is part of the Munich Quantum Valley, supported by the Bavarian state government with funds from the Hightech Agenda Bayern Plus. The National Science Foundation supported JvD in part under PHY-1748958, and GK under Grant No. DMR-1733071.
AG acknowledges support from the Abrahams Postdoctoral Fellowship of the Center for Materials Theory at Rutgers University.

\FloatBarrier
\bibliography{bibfile}

@article{Pelz2026_finiteT,
    author = {Pelz, M. and von Delft, J. and Gleis, A.},
    title = {Dynamical scaling near the pseudogap quantum critical point of the two-dimensional
{H}ubbard model},
    journal = {To be published},
    year = {2026}
}

@article{Pelz2026_Linterp,
      title={Liouvillian interpolation of the self-energy of cluster dynamical mean-field theories}, 
      author={Pelz, M. and von Delft, J. and Gleis, A.},
      year={2026},
      eprint={2602.16351},
      archivePrefix={arXiv},
      primaryClass={cond-mat.str-el},
      journal = {arXiv:2602.16351}
}

@article{Hubbard1963_model,
  author = {Hubbard, J.},
  year = {1963},
  title = {Electron correlations in narrow energy bands},
  journal = {Proc. R. Soc. Lond. A},
  volume = {276},
  pages = {238–257},
  year = {1963},
  doi = {http://doi.org/10.1098/rspa.1963.0204}
}

@article{Weichselbaum2012a_QSpace,
  title = {Non-{A}belian symmetries in tensor networks: {A} quantum symmetry space approach},
  journal = {Annals of Physics},
  volume = {327},
  number = {12},
  pages = {2972-3047},
  year = {2012},
  issn = {0003-4916},
  doi = {https://doi.org/10.1016/j.aop.2012.07.009},
  url = {https://www.sciencedirect.com/science/article/pii/S0003491612001121},
  author = {Andreas Weichselbaum},
}

@article{Weichselbaum2012b_QSpace,
  title = {Tensor networks and the numerical renormalization group},
  author = {Weichselbaum, Andreas},
  journal = {Phys. Rev. B},
  volume = {86},
  issue = {24},
  pages = {245124},
  numpages = {17},
  year = {2012},
  month = {Dec},
  publisher = {American Physical Society},
  doi = {10.1103/PhysRevB.86.245124},
  url = {https://link.aps.org/doi/10.1103/PhysRevB.86.245124}
}

@article{Weichselbaum2020_QSpace,
  title = {{X}-symbols for non-{A}belian symmetries in tensor networks},
  author = {Weichselbaum, Andreas},
  journal = {Phys. Rev. Res.},
  volume = {2},
  issue = {2},
  pages = {023385},
  numpages = {16},
  year = {2020},
  month = {Jun},
  publisher = {American Physical Society},
  doi = {10.1103/PhysRevResearch.2.023385},
  url = {https://link.aps.org/doi/10.1103/PhysRevResearch.2.023385}
}

@Article{Weichselbaum2024_QSpace,
  title={{QSpace - An open-source tensor library for Abelian and non-Abelian symmetries}},
  author={Andreas Weichselbaum},
  journal={SciPost Phys. Codebases},
  pages={40},
  year={2024},
  publisher={SciPost},
  doi={10.21468/SciPostPhysCodeb.40},
  url={https://scipost.org/10.21468/SciPostPhysCodeb.40},
}

@Article{Weichselbaum2024_QSpaceCode,
  title={{Codebase release 4.0 for QSpace}},
  author={Andreas Weichselbaum},
  journal={SciPost Phys. Codebases},
  pages={40-r4.0},
  year={2024},
  publisher={SciPost},
  doi={10.21468/SciPostPhysCodeb.40-r4.0},
  url={https://scipost.org/10.21468/SciPostPhysCodeb.40-r4.0},
}

@Article{Wilson1975_NRG,
  Title = {The renormalization group: {Critical} phenomena and the {Kondo} problem},
  Author = {Wilson, Kenneth G.},
  Journal = {Rev. Mod. Phys.},
  Year = {1975},
  Month = {Oct},
  Pages = {773--840},
  Volume = {47},
  Doi = {10.1103/RevModPhys.47.773},
  Issue = {4},
  Numpages = {0},
  Publisher = {American Physical Society}
}

@article{Krishna-murthy1980_NRG,
  title = {Renormalization-group approach to the {A}nderson model of dilute magnetic alloys. {I}. {S}tatic properties for the symmetric case},
  author = {Krishna-murthy, H. R. and Wilkins, J. W. and Wilson, K. G.},
  journal = {Phys. Rev. B},
  volume = {21},
  issue = {3},
  pages = {1003--1043},
  numpages = {0},
  year = {1980},
  month = {Feb},
  publisher = {American Physical Society},
  doi = {10.1103/PhysRevB.21.1003},
  url = {https://link.aps.org/doi/10.1103/PhysRevB.21.1003}
}

@Article{Anders2005_NRG,
  Title = {Real-Time Dynamics in Quantum-Impurity Systems: A Time-Dependent Numerical Renormalization-Group Approach},
  Author = {Anders, Frithjof B. and Schiller, Avraham},
  Journal = {Phys. Rev. Lett.},
  Year = {2005},
  Month = {Oct},
  Pages = {196801},
  Volume = {95},
  Doi = {10.1103/PhysRevLett.95.196801},
  Issue = {19},
  Numpages = {4},
  Publisher = {American Physical Society}
}

@Article{Bulla2008_NRG,
  Title = {Numerical renormalization group method for quantum impurity systems},
  Author = {Bulla, Ralf and Costi, Theo A. and Pruschke, Thomas},
  Journal = {Rev. Mod. Phys.},
  Year = {2008},
  Month = {Apr},
  Pages = {395--450},
  Volume = {80},
  Doi = {10.1103/RevModPhys.80.395},
  Issue = {2},
  Numpages = {0},
  Publisher = {American Physical Society}
}

@Article{Lee2016_NRG,
  Title = {Adaptive broadening to improve spectral resolution in the numerical renormalization group},
  Author = {Lee, Seung-Sup B. and Weichselbaum, Andreas},
  Journal = {Phys. Rev. B},
  Year = {2016},
  Month = {Dec},
  Pages = {235127},
  Volume = {94},
  Author+an = {1=SSBL},
  Doi = {10.1103/PhysRevB.94.235127},
  Issue = {23},
  Numpages = {15},
  Publisher = {American Physical Society},
  Url = {http://link.aps.org/doi/10.1103/PhysRevB.94.235127}
}

@article{Stadler2016_iNRG,
  title = {Interleaved numerical renormalization group as an efficient multiband impurity solver},
  author = {Stadler, K. M. and Mitchell, A. K. and von Delft, J. and Weichselbaum, A.},
  journal = {Phys. Rev. B},
  volume = {93},
  issue = {23},
  pages = {235101},
  numpages = {16},
  year = {2016},
  month = {Jun},
  publisher = {American Physical Society},
  doi = {10.1103/PhysRevB.93.235101},
  url = {https://link.aps.org/doi/10.1103/PhysRevB.93.235101}
}

@Article{Kugler2022_SEtrick,
  Title = {Improved estimator for numerical renormalization group calculations of the self-energy},
  Author = {Kugler, Fabian B.},
  Journal = {Phys. Rev. B},
  Year = {2022},
  Month = {June},
  Pages = {245132},
  Volume = {105},
  Doi = {10.1103/PhysRevB.105.245132},
  Issue = {24},
  Numpages = {11},
  Publisher = {American Physical Society},
  Url = {https://link.aps.org/doi/10.1103/PhysRevB.105.245132}
}

@article{Georges1996_DMFT,
  title = {Dynamical mean-field theory of strongly correlated fermion systems and the limit of infinite dimensions},
  author = {Georges, Antoine and Kotliar, Gabriel and Krauth, Werner and Rozenberg, Marcelo J.},
  journal = {Rev. Mod. Phys.},
  volume = {68},
  issue = {1},
  pages = {13--125},
  numpages = {0},
  year = {1996},
  month = {Jan},
  publisher = {American Physical Society},
  doi = {10.1103/RevModPhys.68.13},
  url = {https://link.aps.org/doi/10.1103/RevModPhys.68.13}
}

@article{Hettler1998_DCA,
  title = {Nonlocal dynamical correlations of strongly interacting electron systems},
  author = {Hettler, M. H. and Tahvildar-Zadeh, A. N. and Jarrell, M. and Pruschke, T. and Krishnamurthy, H. R.},
  journal = {Phys. Rev. B},
  Year = {1998},
  Month = {Sep},
  Pages = {R7475--R7479},
  Volume = {58},
  Doi = {10.1103/PhysRevB.58.R7475},
  Issue = {12},
  Numpages = {0},
  Publisher = {American Physical Society},
  Url = {https://link.aps.org/doi/10.1103/PhysRevB.58.R7475}
}

@Article{Hettler2000_DCA,
  Title = {Dynamical cluster approximation: {N}onlocal dynamics of correlated electron systems},
  Author = {Hettler, M. H. and Mukherjee, M. and Jarrell, M. and Krishnamurthy, H. R.},
  Journal = {Phys. Rev. B},
  Year = {2000},
  Month = {May},
  Pages = {12739--12756},
  Volume = {61},
  Doi = {10.1103/PhysRevB.61.12739},
  Issue = {19},
  Numpages = {0},
  Publisher = {American Physical Society},
  Url = {https://link.aps.org/doi/10.1103/PhysRevB.61.12739}
}

@article{Maier2005_clusters,
  title = {Quantum cluster theories},
  author = {Maier, Thomas and Jarrell, Mark and Pruschke, Thomas and Hettler, Matthias H.},
  journal = {Rev. Mod. Phys.},
  volume = {77},
  issue = {3},
  pages = {1027--1080},
  numpages = {0},
  year = {2005},
  month = {Oct},
  publisher = {American Physical Society},
  doi = {10.1103/RevModPhys.77.1027},
  url = {https://link.aps.org/doi/10.1103/RevModPhys.77.1027}
}

@article{Gull2010_patching-pg,
  title = {Momentum-space anisotropy and pseudogaps: A comparative cluster dynamical mean-field analysis of the doping-driven metal-insulator transition in the two-dimensional {H}ubbard model},
  author = {Gull, E. and Ferrero, M. and Parcollet, O. and Georges, A. and Millis, A. J.},
  journal = {Phys. Rev. B},
  volume = {82},
  issue = {15},
  pages = {155101},
  numpages = {14},
  year = {2010},
  month = {Oct},
  publisher = {American Physical Society},
  doi = {10.1103/PhysRevB.82.155101},
  url = {https://link.aps.org/doi/10.1103/PhysRevB.82.155101}
}

@InCollection{Potthoff2016_cDMFT,
  author    = {Potthoff, M.},
  booktitle = {{DMFT: From Infinite Dimensions to Real Materials}},
  series    = {Modeling and Simulation},
  volume    = {8},
  pages     = {5.1--5.33},
  publisher = {Forschungszentrum J{\"u}lich},
  title     = {Cluster Extensions of Dynamical Mean-Field Theory},
  editor    = {Pavarini, E. and Koch, E. and Lichtenstein, A. and Vollhardt, D.},
  year      = {2018},
  isbn      = {978-3-95806-313-6},
  URL       = {https://www.cond-mat.de/events/correl18/manuscripts/potthoff.pdf}
}

@article{Keimer2015,
  author = {Keimer, B. and Kivelson, S. A. and Norman, M. R. and Uchida, S. and Zaanen, J.},
  title = {From quantum matter to high-temperature superconductivity in copper oxides},
  journal = {Nature},
  volume = {518},
  pages = {179--186},
  year = {2015},
  doi = {10.1038/nature14165}
}

@article{Badoux2016,
  author = {Badoux, S. and Tabis, W. and Laliberté, F. and Grissonnanche, G. and Vignolle, B. and Vignolles, D. and Béard, J. and Bonn, D. A. and Hardy, W. N. and Liang, R. and Doiron-Leyraud, N. and Taillefer, L. and Proust, C.},
  title = {Change of carrier density at the pseudogap critical point of a cuprate superconductor},
  journal = {Nature},
  volume = {531},
  pages = {210--214},
  year = {2016},
  doi = {10.1038/nature16983}
}

@article{DoironLeyraud2007,
  author = {Doiron-Leyraud, N. and Proust, C. and LeBoeuf, D. and Levallois, J. and Bonnemaison, J. B. and Liang, R. and Bonn, D. A. and Hardy, W. N. and Taillefer, L.},
  title = {Quantum oscillations and the {F}ermi surface in an underdoped high-{$T_c$} superconductor},
  journal = {Nature},
  volume = {447},
  pages = {565--568},
  year = {2007},
  doi = {10.1038/nature05872}
}

@article{Sebastian2012,
  author = {Sebastian, S. E. and Harrison, N. and Lonzarich, G. G.},
  title = {Towards resolution of the {F}ermi surface in underdoped {high-$T_c$} superconductors},
  journal = {Reports on Progress in Physics},
  volume = {75},
  pages = {102501},
  year = {2012},
  doi = {10.1088/0034-4885/75/10/102501}
}

@article{Vishik2012,
  author = {Vishik, I. M. and Hashimoto, M. and He, R.-H. and Lee, W.-S. and Schmitt, F. T. and Lu, D. H. and Moore, R. G. and Zhang, C. and Meevasana, W. and Sasagawa, T. and Uchida, S. and Fujita, K. and Ishida, S. and Ishikado, M. and Yoshida, Y. and Eisaki, H. and Hussain, Z. and Devereaux, T. P. and Shen, Z.-X.},
  title = {Phase competition in trisected superconducting dome},
  journal = {Proceedings of the National Academy of Sciences},
  volume = {109},
  pages = {18332--18337},
  year = {2012},
  doi = {10.1073/pnas.1209471109}
}

@article{Varma1997,
  author = {Varma, C. M.},
  title = {{Non-Fermi-liquid} states and pairing instability of a general model of copper oxide metals},
  journal = {Physical Review B},
  volume = {55},
  pages = {14554--14580},
  year = {1997},
  doi = {10.1103/PhysRevB.55.14554}
}

@article{Varma1999,
  author = {Varma, C. M.},
  title = {Pseudogap phase and the quantum-critical point in copper-oxide metals},
  journal = {Physical Review Letters},
  volume = {83},
  pages = {3538--3541},
  year = {1999},
  doi = {10.1103/PhysRevLett.83.3538}
}

@article{Millis1993,
  author = {Millis, A. J.},
  title = {Effect of a nonzero temperature on quantum critical points in itinerant fermion systems},
  journal = {Physical Review B},
  volume = {48},
  pages = {7183--7196},
  year = {1993},
  doi = {10.1103/PhysRevB.48.7183}
}

@article{Senthil2004,
  author = {Senthil, T. and Vojta, M. and Sachdev, S.},
  title = {Weak magnetism and non-{F}ermi liquids near heavy-fermion critical points},
  journal = {Physical Review B},
  volume = {69},
  pages = {035111},
  year = {2004},
  doi = {10.1103/PhysRevB.69.035111}
}

@article{Damascelli2003,
  author = {Damascelli, A. and Hussain, Z. and Shen, Z.-X.},
  title = {Angle-resolved photoemission studies of the cuprate superconductors},
  journal = {Reviews of Modern Physics},
  volume = {75},
  pages = {473--541},
  year = {2003},
  doi = {10.1103/RevModPhys.75.473}
}

@article{Norman1998,
  author = {Norman, M. R. and Ding, H. and Randeria, M. and Campuzano, J. C. and Yokoya, T. and Takeuchi, T. and Takahashi, T. and Mochiku, T. and Kadowaki, K. and Guptasarma, P. and Hinks, D. G.},
  title = {Destruction of the {F}ermi surface in underdoped {high-$T_c$} superconductors},
  journal = {Nature},
  volume = {392},
  pages = {157--160},
  year = {1998},
  doi = {10.1038/32366}
}

@article{He2011,
  author = {He, R.-H. and Hashimoto, M. and Karapetyan, H. and Koralek, J. D. and Hinton, J. P. and Testaud, J. P. and Nathan, V. and Yoshida, Y. and Yao, H. and Tanaka, K. and Meevasana, W. and Moore, R. G. and Lu, D. H. and Mo, S.-K. and Ishikado, M. and Eisaki, H. and Hussain, Z. and Devereaux, T. P. and Kivelson, S. A. and Orenstein, J. and Kapitulnik, A. and Shen, Z.-X.},
  title = {From a Single-Band Metal to a High-Temperature Superconductor via Two Thermal Phase Transitions},
  journal = {Science},
  volume = {331},
  pages = {1579--1583},
  year = {2011},
  doi = {10.1126/science.1198415}
}

@article{Chen2019,
author = {Chen, Su-Di  and Hashimoto, Makoto  and He, Yu and Song, Dongjoon and Xu, Ke-Jun and He, Jun-Feng and Thomas P. Devereaux  and Hiroshi, Eisaki  and Lu, Dong-Hui and Zaanen, Jan and Shen, Zhi-Xun},
title = {Incoherent strange metal sharply bounded by a critical doping in {B}i2212},
journal = {Science},
volume = {366},
number = {6469},
pages = {1099-1102},
year = {2019},
doi = {10.1126/science.aaw8850}
}

@article{Reber2019,
author = {Reber, T. J. and Zhou, X. and Plumb, N. C. and Parham, S. and Waugh, J. A. and Cao, Y. and Sun, Z. and Li, H. and Wang, Q. and Wen, J. S. and Xu, Z. J. and Gu, G. and Yoshida, Y. and Eisaki, H. and Arnold, G. B. and Dessau, D. S.},
title = {A unified form of low-energy nodal electronic interactions in hole-doped cuprate superconductors},
journal = {Nat. Commun.},
volume ={10},
pages = {5737},
year = {2019},
doi={10.1038/s41467-019-13497-4}
}

@article{Michon2019,
  author = {Michon, B. and Girod, C. and Badoux, S. and Ka{\v{c}}mar{\v{c}}{\'i}k, J. and Ma, Q. and Dragomir, M. and Dabkowska, H. A. and Gaulin, B. D. and Zhou, J.-S. and Pyon, S. and Takayama, T. and Takagi, H. and Verret, S. and Doiron-Leyraud, N. and Marcenat, C. and Taillefer, L. and Klein, T.},
  title = {Thermodynamic signatures of quantum criticality in cuprate superconductors},
  journal = {Nature},
  volume = {567},
  pages = {218--222},
  year = {2019},
  doi = {10.1038/s41586-019-0932-x}
}

@article{Varma1989_MFL,
  author = {Varma, C. M. and Littlewood, P. B. and Schmitt-Rink, S. and Abrahams, E. and Ruckenstein, A. E.},
  title = {Phenomenology of the normal state of {Cu-O} high-temperature superconductors},
  journal = {Phys. Rev. Lett.},
  volume = {63},
  pages = {1996--1999},
  year = {1989},
  doi = {10.1103/PhysRevLett.63.1996}
}

@article{Sordi2010,
  author = {Sordi, G. and Haule, K. and Tremblay, A.-M. S.},
  title = {Finite Doping Signatures of the {Mott} Transition in the Two-Dimensional {Hubbard} Model},
  journal = {Phys. Rev. Lett.},
  volume = {104},
  pages = {226402},
  year = {2010},
  doi = {10.1103/PhysRevLett.104.226402}
}

@article{Sordi2012_Widom,
  author = {Sordi, G. and S{\'e}mon, P. and Haule, K. and Tremblay, A.-M. S.},
  title = {Pseudogap temperature as a {Widom} line in doped {Mott} insulators},
  journal = {Scientific Reports},
  volume = {2},
  pages = {547},
  year = {2012},
  doi = {10.1038/srep00547}
}

@article{Sordi2012_PRL,
  author = {Sordi, G. and S{\'e}mon, P. and Haule, K. and Tremblay, A.-M. S.},
  title = {Strong Coupling Superconductivity, Pseudogap, and {Mott} Transition},
  journal = {Phys. Rev. Lett.},
  volume = {108},
  pages = {216401},
  year = {2012},
  doi = {10.1103/PhysRevLett.108.216401}
}

@article{Vidhyadhiraja2009,
  author = {Vidhyadhiraja, N. S. and Macridin, A. and Sen, C. and Jarrell, M. and Ma, Michael},
  title = {Quantum Critical Point at Finite Doping in the 2D {Hubbard} Model: A Dynamical Cluster Quantum {Monte Carlo} Study},
  journal = {Phys. Rev. Lett.},
  volume = {102},
  pages = {206407},
  year = {2009},
  doi = {10.1103/PhysRevLett.102.206407}
}

@article{Civelli2005_FermiSurfaceBreakup,
  author = {Civelli, M. and Capone, M. and Kancharla, S. S. and Parcollet, O. and Kotliar, G.},
  title = {Dynamical Breakup of the {F}ermi Surface in a Doped {M}ott Insulator},
  journal = {Phys. Rev. Lett.},
  volume = {95},
  pages = {106402},
  year = {2005},
  doi = {10.1103/PhysRevLett.95.106402}
}

@article{Stanescu2006_FermiArcs,
  author = {Stanescu, Tudor D. and Kotliar, Gabriel},
  title = {{F}ermi arcs and hidden zeros of the {G}reen function in the pseudogap state},
  journal = {Phys. Rev. B},
  volume = {74},
  pages = {125110},
  year = {2006},
  doi = {10.1103/PhysRevB.74.125110}
}

@article{Ferrero2009_FermiArcs,
  author = {Ferrero, M. and Cornaglia, P. S. and De Leo, L. and Parcollet, O. and Kotliar, G. and Georges, A.},
  title = {Pseudogap opening and formation of {F}ermi arcs as an orbital-selective {M}ott transition in momentum space},
  journal = {Phys. Rev. B},
  volume = {80},
  pages = {064501},
  year = {2009},
  doi = {10.1103/PhysRevB.80.064501}
}

@article{Hertz1976,
  author = {Hertz, J. A.},
  title = {Quantum critical phenomena},
  journal = {Phys. Rev. B},
  volume = {14},
  pages = {1165--1184},
  year = {1976},
  doi = {10.1103/PhysRevB.14.1165}
}

@book{Moriya1985,
  author = {Moriya, Toru},
  title = {Spin Fluctuations in Itinerant Electron Magnetism},
  publisher = {Springer},
  address = {Berlin},
  series = {Springer Series in Solid-State Sciences},
  volume = {56},
  year = {1985},
  doi = {10.1007/978-3-642-82499-9}
}

@article{Collignon2017_PRB,
  author = {Collignon, C. and Badoux, S. and Afshar, S. A. A. and Michon, B. and Laliberté, F. and Cyr-Choinière, O. and Zhou, J.-S. and Licciardello, S. and Wiedmann, S. and Doiron-Leyraud, N. and Taillefer, L.},
  title = {Fermi-surface transformation across the pseudogap critical point of the cuprate superconductor {La$_{1.6-x}$Nd$_{0.4}$Sr$_x$CuO$_4$}},
  journal = {Phys. Rev. B},
  volume = {95},
  pages = {224517},
  year = {2017},
  doi = {10.1103/PhysRevB.95.224517}
}

@article{Fang2022_ADMR,
  author = {Fang, Yawen and Grissonnanche, Gaël and Legros, Anaëlle and Verret, Simon and Laliberté, Francis and Collignon, Clément and Ataei, Amirreza and Dion, Maxime and Zhou, Jianshi and Graf, David and Lawler, M. J. and Goddard, Paul and Taillefer, Louis and Ramshaw, B. J.},
  title = {Fermi surface transformation at the pseudogap critical point of a cuprate superconductor},
  journal = {Nature Physics},
  volume = {18},
  pages = {558--564},
  year = {2022},
  doi = {10.1038/s41567-022-01514-1}
}

@article{Sachdev2010,
author = {Sachdev, Subir},
title = {Where is the quantum critical point in the cuprate superconductors?},
journal = {physica status solidi (b)},
volume = {247},
number = {3},
pages = {537-543},
doi = {https://doi.org/10.1002/pssb.200983037},
year = {2010}
}

@article{supplement,
journal = {See the Supplemental Material at [url] for additional information on the model and methods; discussion about susceptiblities and extraction of energy scales; doping and patch momentum dependence of the Green's function, the self energy, and the hybridization; cluster data corresponding to the $\mc{L}$-interpolated spectral function}
}

@Article{Kanigel2006,
author={Kanigel, A.
and Norman, M. R.
and Randeria, M.
and Chatterjee, U.
and Souma, S.
and Kaminski, A.
and Fretwell, H. M.
and Rosenkranz, S.
and Shi, M.
and Sato, T.
and Takahashi, T.
and Li, Z. Z.
and Raffy, H.
and Kadowaki, K.
and Hinks, D.
and Ozyuzer, L.
and Campuzano, J. C.},
title={Evolution of the pseudogap from Fermi arcs to the nodal liquid},
journal={Nature Physics},
year={2006},
month={Jul},
day={01},
volume={2},
number={7},
pages={447-451},
issn={1745-2481},
doi={10.1038/nphys334},
url={https://doi.org/10.1038/nphys334}
}

@article{Timusk1999_Pseudogap,
  author = {Timusk, T. and Statt, B.},
  title = {The pseudogap in high-temperature superconductors: an experimental survey},
  journal = {Reports on Progress in Physics},
  volume = {62},
  number = {1},
  pages = {61--122},
  year = {1999},
  doi = {10.1088/0034-4885/62/1/002}
}

@article{Lee2006,
  title = {Doping a Mott insulator: Physics of high-temperature superconductivity},
  author = {Lee, Patrick A. and Nagaosa, Naoto and Wen, Xiao-Gang},
  journal = {Rev. Mod. Phys.},
  volume = {78},
  issue = {1},
  pages = {17--85},
  numpages = {0},
  year = {2006},
  month = {Jan},
  publisher = {American Physical Society},
  doi = {10.1103/RevModPhys.78.17},
  url = {https://link.aps.org/doi/10.1103/RevModPhys.78.17}
}

@article{Huscroft2001,
  title = {Pseudogaps in the 2D Hubbard Model},
  author = {Huscroft, C. and Jarrell, M. and Maier, Th. and Moukouri, S. and Tahvildarzadeh, A. N.},
  journal = {Phys. Rev. Lett.},
  volume = {86},
  issue = {1},
  pages = {139--142},
  numpages = {0},
  year = {2001},
  month = {Jan},
  publisher = {American Physical Society},
  doi = {10.1103/PhysRevLett.86.139},
  url = {https://link.aps.org/doi/10.1103/PhysRevLett.86.139}
}

@article{Kyung2006,
  title = {Pseudogap induced by short-range spin correlations in a doped Mott insulator},
  author = {Kyung, B. and Kancharla, S. S. and S\'en\'echal, D. and Tremblay, A.-M. S. and Civelli, M. and Kotliar, G.},
  journal = {Phys. Rev. B},
  volume = {73},
  issue = {16},
  pages = {165114},
  numpages = {6},
  year = {2006},
  month = {Apr},
  publisher = {American Physical Society},
  doi = {10.1103/PhysRevB.73.165114},
  url = {https://link.aps.org/doi/10.1103/PhysRevB.73.165114}
}

@article{Macridin2006,
  title = {Pseudogap and Antiferromagnetic Correlations in the Hubbard Model},
  author = {Macridin, Alexandru and Jarrell, M. and Maier, Thomas and Kent, P. R. C. and D'Azevedo, Eduardo},
  journal = {Phys. Rev. Lett.},
  volume = {97},
  issue = {3},
  pages = {036401},
  numpages = {4},
  year = {2006},
  month = {Jul},
  publisher = {American Physical Society},
  doi = {10.1103/PhysRevLett.97.036401},
  url = {https://link.aps.org/doi/10.1103/PhysRevLett.97.036401}
}

@article{Werner2009,
  title = {Momentum-sector-selective metal-insulator transition in the eight-site dynamical mean-field approximation to the Hubbard model in two dimensions},
  author = {Gull, Emanuel and Parcollet, Olivier and Werner, Philipp and Millis, Andrew J.},
  journal = {Phys. Rev. B},
  volume = {80},
  issue = {24},
  pages = {245102},
  numpages = {13},
  year = {2009},
  month = {Dec},
  publisher = {American Physical Society},
  doi = {10.1103/PhysRevB.80.245102},
  url = {https://link.aps.org/doi/10.1103/PhysRevB.80.245102}
}

@article{Liebsch2009,
  title = {Finite-temperature exact diagonalization cluster dynamical mean-field study of the two-dimensional Hubbard model: Pseudogap, non-Fermi-liquid behavior, and particle-hole asymmetry},
  author = {Liebsch, Ansgar and Tong, Ning-Hua},
  journal = {Phys. Rev. B},
  volume = {80},
  issue = {16},
  pages = {165126},
  numpages = {18},
  year = {2009},
  month = {Oct},
  publisher = {American Physical Society},
  doi = {10.1103/PhysRevB.80.165126},
  url = {https://link.aps.org/doi/10.1103/PhysRevB.80.165126}
}

@article{Mikelsons2009,
  title = {Thermodynamics of the quantum critical point at finite doping in the two-dimensional Hubbard model studied via the dynamical cluster approximation},
  author = {Mikelsons, K. and Khatami, E. and Galanakis, D. and Macridin, A. and Moreno, J. and Jarrell, M.},
  journal = {Phys. Rev. B},
  volume = {80},
  issue = {14},
  pages = {140505(R)},
  numpages = {4},
  year = {2009},
  month = {Oct},
  publisher = {American Physical Society},
  doi = {10.1103/PhysRevB.80.140505},
  url = {https://link.aps.org/doi/10.1103/PhysRevB.80.140505}
}

@article{Khatami2010,
  title = {Quantum criticality due to incipient phase separation in the two-dimensional Hubbard model},
  author = {Khatami, E. and Mikelsons, K. and Galanakis, D. and Macridin, A. and Moreno, J. and Scalettar, R. T. and Jarrell, M.},
  journal = {Phys. Rev. B},
  volume = {81},
  issue = {20},
  pages = {201101(R)},
  numpages = {4},
  year = {2010},
  month = {May},
  publisher = {American Physical Society},
  doi = {10.1103/PhysRevB.81.201101},
  url = {https://link.aps.org/doi/10.1103/PhysRevB.81.201101}
}

@article{Yang2011,
  title = {Proximity of the Superconducting Dome and the Quantum Critical Point in the Two-Dimensional Hubbard Model},
  author = {Yang, S.-X. and Fotso, H. and Su, S.-Q. and Galanakis, D. and Khatami, E. and She, J.-H. and Moreno, J. and Zaanen, J. and Jarrell, M.},
  journal = {Phys. Rev. Lett.},
  volume = {106},
  issue = {4},
  pages = {047004},
  numpages = {4},
  year = {2011},
  month = {Jan},
  publisher = {American Physical Society},
  doi = {10.1103/PhysRevLett.106.047004},
  url = {https://link.aps.org/doi/10.1103/PhysRevLett.106.047004}
}

@article{Sordi2011,
  title = {Mott physics and first-order transition between two metals in the normal-state phase diagram of the two-dimensional Hubbard model},
  author = {Sordi, G. and Haule, K. and Tremblay, A.-M. S.},
  journal = {Phys. Rev. B},
  volume = {84},
  issue = {7},
  pages = {075161},
  numpages = {25},
  year = {2011},
  month = {Aug},
  publisher = {American Physical Society},
  doi = {10.1103/PhysRevB.84.075161},
  url = {https://link.aps.org/doi/10.1103/PhysRevB.84.075161}
}

@article{Sordi2013,
  title = {$c$-axis resistivity, pseudogap, superconductivity, and Widom line in doped Mott insulators},
  author = {Sordi, G. and S\'emon, P. and Haule, K. and Tremblay, A.-M. S.},
  journal = {Phys. Rev. B},
  volume = {87},
  issue = {4},
  pages = {041101(R)},
  numpages = {5},
  year = {2013},
  month = {Jan},
  publisher = {American Physical Society},
  doi = {10.1103/PhysRevB.87.041101},
  url = {https://link.aps.org/doi/10.1103/PhysRevB.87.041101}
}

@article{Sordi2019,
  title = {Specific heat maximum as a signature of Mott physics in the two-dimensional Hubbard model},
  author = {Sordi, G. and Walsh, C. and S\'emon, P. and Tremblay, A.-M. S.},
  journal = {Phys. Rev. B},
  volume = {100},
  issue = {12},
  pages = {121105(R)},
  numpages = {6},
  year = {2019},
  month = {Sep},
  publisher = {American Physical Society},
  doi = {10.1103/PhysRevB.100.121105},
  url = {https://link.aps.org/doi/10.1103/PhysRevB.100.121105}
}

@Article{Meixner2024,
	title={{Mott transition and pseudogap of the square-lattice Hubbard model: Results from center-focused cellular dynamical mean-field theory}},
	author={Michael Meixner and Henri Menke and Marcel Klett and Sarah Heinzelmann and Sabine Andergassen and Philipp Hansmann and Thomas Schäfer},
	journal={SciPost Phys.},
	volume={16},
	pages={059},
	year={2024},
	publisher={SciPost},
	doi={10.21468/SciPostPhys.16.2.059},
	url={https://scipost.org/10.21468/SciPostPhys.16.2.059},
}

@article{Scheurer2018,
author = {Mathias S. Scheurer  and Shubhayu Chatterjee  and Wei Wu  and Michel Ferrero  and Antoine Georges  and Subir Sachdev },
title = {Topological order in the pseudogap metal},
journal = {Proceedings of the National Academy of Sciences},
volume = {115},
number = {16},
pages = {E3665-E3672},
year = {2018},
doi = {10.1073/pnas.1720580115},
URL = {https://www.pnas.org/doi/abs/10.1073/pnas.1720580115}
}

@article{Wu2018,
  title = {Pseudogap and Fermi-Surface Topology in the Two-Dimensional Hubbard Model},
  author = {Wu, Wei and Scheurer, Mathias S. and Chatterjee, Shubhayu and Sachdev, Subir and Georges, Antoine and Ferrero, Michel},
  journal = {Phys. Rev. X},
  volume = {8},
  issue = {2},
  pages = {021048},
  numpages = {17},
  year = {2018},
  month = {May},
  publisher = {American Physical Society},
  doi = {10.1103/PhysRevX.8.021048},
  url = {https://link.aps.org/doi/10.1103/PhysRevX.8.021048}
}

@article{Mitchell2014_iNRG,
  title = {Generalized Wilson chain for solving multichannel quantum impurity problems},
  author = {Mitchell, A. K. and Galpin, M. R. and Wilson-Fletcher, S. and Logan, D. E. and Bulla, R.},
  journal = {Phys. Rev. B},
  volume = {89},
  issue = {12},
  pages = {121105},
  numpages = {5},
  year = {2014},
  month = {Mar},
  publisher = {American Physical Society},
  doi = {10.1103/PhysRevB.89.121105},
  url = {https://link.aps.org/doi/10.1103/PhysRevB.89.121105}
}

@article{Qin2022_HM,
   author = "Qin, Mingpu and Schäfer, Thomas and Andergassen, Sabine and Corboz, Philippe and Gull, Emanuel",
   title = "The {H}ubbard Model: A Computational Perspective",
   journal= "Annual Review of Condensed Matter Physics",
   year = "2022",
   volume = "13",
   number = "Volume 13, 2022",
   pages = "275-302",
   doi = "https://doi.org/10.1146/annurev-conmatphys-090921-033948",
   url = "https://www.annualreviews.org/content/journals/10.1146/annurev-conmatphys-090921-033948",
   publisher = "Annual Reviews",
   issn = "1947-5462",
   type = "Journal Article"
  }

@article{Backes2022_genCavity,
  title = {Nonlocal correlation effects in fermionic many-body systems: Overcoming the noncausality problem},
  author = {Backes, Steffen and Sim, Jae-Hoon and Biermann, Silke},
  journal = {Phys. Rev. B},
  volume = {105},
  issue = {24},
  pages = {245115},
  numpages = {13},
  year = {2022},
  month = {Jun},
  publisher = {American Physical Society},
  doi = {10.1103/PhysRevB.105.245115},
  url = {https://link.aps.org/doi/10.1103/PhysRevB.105.245115}
}

@article{Castellani1995,
  title = {Singular Quasiparticle Scattering in the Proximity of Charge Instabilities},
  author = {Castellani, C. and Di Castro, C. and Grilli, M.},
  journal = {Phys. Rev. Lett.},
  volume = {75},
  issue = {25},
  pages = {4650--4653},
  numpages = {0},
  year = {1995},
  month = {Dec},
  publisher = {American Physical Society},
  doi = {10.1103/PhysRevLett.75.4650},
  url = {https://link.aps.org/doi/10.1103/PhysRevLett.75.4650}
}

@article{Tallon1999,
author = {Tallon, J. L. and Loram, J. W. and Williams, G. V. M. and Cooper, J. R. and Fisher, I. R. and Johnson, J. D. and Staines, M. P. and Bernhard, C.},
title = {Critical Doping in Overdoped High-$T_c$ Superconductors: a Quantum Critical Point?},
journal = {physica status solidi (b)},
volume = {215},
number = {1},
pages = {531-540},
doi = {https://doi.org/10.1002/(SICI)1521-3951(199909)215:1<531::AID-PSSB531>3.0.CO;2-W},
year = {1999}
}

@article{Haule2007,
  title = {Strongly correlated superconductivity: {A} plaquette dynamical mean-field theory study},
  author = {Haule, Kristjan and Kotliar, Gabriel},
  journal = {Phys. Rev. B},
  volume = {76},
  issue = {10},
  pages = {104509},
  numpages = {37},
  year = {2007},
  month = {Sep},
  publisher = {American Physical Society},
  doi = {10.1103/PhysRevB.76.104509},
  url = {https://link.aps.org/doi/10.1103/PhysRevB.76.104509}
}

@article{Tremblay2006_num,
    author = {Tremblay, A.-M. S. and Kyung, B. and Sénéchal, D.},
    title = {Pseudogap and high-temperature superconductivity from weak to strong coupling. Towards a quantitative theory},
    journal = {Low Temperature Physics},
    volume = {32},
    number = {4},
    pages = {424-451},
    year = {2006},
    issn = {1063-777X},
    doi = {10.1063/1.2199446},
    url = {https://doi.org/10.1063/1.2199446}
}

@article{Peterson2010_compressibility,
  title = {Kelvin formula for thermopower},
  author = {Peterson, Michael R. and Shastry, B. Sriram},
  journal = {Phys. Rev. B},
  volume = {82},
  issue = {19},
  pages = {195105},
  numpages = {5},
  year = {2010},
  month = {Nov},
  publisher = {American Physical Society},
  doi = {10.1103/PhysRevB.82.195105},
  url = {https://link.aps.org/doi/10.1103/PhysRevB.82.195105}
}

@article{Reber2012_ARPES_cuprates,
  author  = {Reber, T.J. and Plumb, N.C. and Sun, Z. and Cao, Y. and Wang, Q. and McElroy, K. and Iwasawa, H. and Arita, M. and Wen, J. S. and Xu, Z. J. and Gu, G. and Yoshida, Y. and Eisaki, H. and Aiura, Y. and Dessau, D. S.},
  title = {The origin and non-quasiparticle nature of {F}ermi arcs in $\mr{Bi}_2\mr{Sr}_2\mr{CaCu}_2\mr{O}_{8+\delta}$},
  journal = {Nature Phys.},
  volume = {8},
  pages = {606-610},
  doi = {https://doi.org/10.1038/nphys2352},
  year = {2012}
}

@article{Li2022,
    author = {Li, R. and She, ZS},
    title = {Unified energy law for fluctuating density wave orders in cuprate pseudogap phase},
    journal = {Commun. Phys.},
    year = {2022},
    volume ={5},
    pages = {13},
    doi = {https://doi.org/10.1038/s42005-021-00789-9}
}

@article{Nie2015,
  title = {Fluctuating orders and quenched randomness in the cuprates},
  author = {Nie, Laimei and Sierens, Lauren E. Hayward and Melko, Roger G. and Sachdev, Subir and Kivelson, Steven A.},
  journal = {Phys. Rev. B},
  volume = {92},
  issue = {17},
  pages = {174505},
  numpages = {13},
  year = {2015},
  month = {Nov},
  publisher = {American Physical Society},
  doi = {10.1103/PhysRevB.92.174505},
  url = {https://link.aps.org/doi/10.1103/PhysRevB.92.174505}
}

@article{Hayward2014,
author = {Hayward, Lauren E. and Hawthorn, David G. and Melko, Roger G. and Sachdev, Subir},
title = {Angular Fluctuations of a Multicomponent Order Describe the Pseudogap of {YBa}$_2${Cu}$_3${O}$_{6+x}$},
journal = {Science},
volume = {343},
number = {6177},
pages = {1336-1339},
year = {2014},
doi = {10.1126/science.1246310},
URL = {https://www.science.org/doi/abs/10.1126/science.1246310}
}

@article{Sachdev2004,
  title = {Competing orders in thermally fluctuating superconductors in two dimensions},
  author = {Sachdev, Subir and Demler, Eugene},
  journal = {Phys. Rev. B},
  volume = {69},
  issue = {14},
  pages = {144504},
  numpages = {9},
  year = {2004},
  month = {Apr},
  publisher = {American Physical Society},
  doi = {10.1103/PhysRevB.69.144504},
  url = {https://link.aps.org/doi/10.1103/PhysRevB.69.144504}
}

@article{Bonetti2026,
doi = {10.1088/1361-6633/ae530d},
url = {https://doi.org/10.1088/1361-6633/ae530d},
year = {2026},
month = {apr},
publisher = {IOP Publishing},
volume = {89},
number = {4},
pages = {044501},
author = {Bonetti, Pietro M and Christos, Maine and Nikolaenko, Alexander and Patel, Aavishkar A and Sachdev, Subir},
title = {Fractionalized {F}ermi liquids and the cuprate phase diagram},
journal = {Reports on Progress in Physics}
}

@article{Moon2011,
  title = {Underdoped cuprates as fractionalized {F}ermi liquids: Transition to superconductivity},
  author = {Moon, Eun Gook and Sachdev, Subir},
  journal = {Phys. Rev. B},
  volume = {83},
  issue = {22},
  pages = {224508},
  numpages = {11},
  year = {2011},
  month = {Jun},
  publisher = {American Physical Society},
  doi = {10.1103/PhysRevB.83.224508},
  url = {https://link.aps.org/doi/10.1103/PhysRevB.83.224508}
}

@article{Wu2024,
  title = {Deconfined {F}ermi liquid to {F}ermi liquid transition and superconducting instability},
  author = {Wu, Xiaofan and Yang, Hui and Zhang, Ya-Hui},
  journal = {Phys. Rev. B},
  volume = {110},
  issue = {12},
  pages = {125122},
  numpages = {18},
  year = {2024},
  month = {Sep},
  publisher = {American Physical Society},
  doi = {10.1103/PhysRevB.110.125122},
  url = {https://link.aps.org/doi/10.1103/PhysRevB.110.125122}
}

@article{Wang2022,
author = {Jiangfan Wang  and Yung-Yeh Chang  and Chung-Hou Chung },
title = {A mechanism for the strange metal phase in rare-earth intermetallic compounds},
journal = {Proceedings of the National Academy of Sciences},
volume = {119},
number = {10},
pages = {e2116980119},
year = {2022},
doi = {10.1073/pnas.2116980119}
}

@article{Senthil2003,
  title = {Fractionalized Fermi Liquids},
  author = {Senthil, T. and Sachdev, Subir and Vojta, Matthias},
  journal = {Phys. Rev. Lett.},
  volume = {90},
  issue = {21},
  pages = {216403},
  numpages = {4},
  year = {2003},
  month = {May},
  publisher = {American Physical Society},
  doi = {10.1103/PhysRevLett.90.216403},
  url = {https://link.aps.org/doi/10.1103/PhysRevLett.90.216403}
}

@article{Lee2017,
  title = {Doublon-Holon Origin of the Subpeaks at the Hubbard Band Edges},
  author = {Lee, Seung-Sup B. and von Delft, Jan and Weichselbaum, Andreas},
  journal = {Phys. Rev. Lett.},
  volume = {119},
  issue = {23},
  pages = {236402},
  numpages = {5},
  year = {2017},
  month = {Dec},
  publisher = {American Physical Society},
  doi = {10.1103/PhysRevLett.119.236402},
  url = {https://link.aps.org/doi/10.1103/PhysRevLett.119.236402}
}

@Article{Eberlein2016,
  author  = {Eberlein, Andreas and Metzner, Walter and Sachdev, Subir and Yamase, Hiroyuki},
  journal = {Phys. Rev. Lett.},
  title   = {{Fermi} surface reconstruction and drop in the {Hall} number due to spiral antiferromagnetism in high-${T}_{c}$ cuprates},
  year    = {2016},
  number  = {18},
  pages   = {187001},
  volume  = {117},
  doi     = {10.1103/PhysRevLett.117.187001}
}

@Article{Verret2017,
  author  = {Verret, S. and Simard, O. and Charlebois, M. and S\'en\'echal, D. and Tremblay, A.-M. S.},
  journal = {Phys. Rev. B},
  title   = {Phenomenological theories of the low-temperature pseudogap: {Hall} number, specific heat, and {Seebeck} coefficient},
  year    = {2017},
  number  = {12},
  pages   = {125139},
  volume  = {96},
  doi     = {10.1103/PhysRevB.96.125139}
}

@Article{Bonetti2020,
  author  = {Bonetti, Pietro M. and Mitscherling, Johannes and Vilardi, Demetrio and Metzner, Walter},
  journal = {Phys. Rev. B},
  title   = {Charge carrier drop at the onset of pseudogap behavior in the two-dimensional {Hubbard} model},
  year    = {2020},
  number  = {16},
  pages   = {165142},
  volume  = {101},
  doi     = {10.1103/PhysRevB.101.165142}
}

@Article{Bonetti2022,
  author  = {Bonetti, Pietro M. and Metzner, Walter},
  journal = {Phys. Rev. B},
  title   = {{SU(2)} gauge theory of the pseudogap phase in the two-dimensional {Hubbard} model},
  year    = {2022},
  number  = {20},
  pages   = {205152},
  volume  = {106},
  doi     = {10.1103/PhysRevB.106.205152}
}

@Article{Forni2026,
  author  = {Forni, Paulo and Bonetti, Pietro M. and M\"uller-Groeling, Henrik and Vilardi, Demetrio and Metzner, Walter},
  journal = {Phys. Rev. B},
  title   = {Spin susceptibility in a pseudogap state with fluctuating spiral magnetic order},
  year    = {2026},
  number  = {4},
  pages   = {045144},
  volume  = {113},
  doi     = {10.1103/zm7b-jdzf}
}

@Article{Klett2022,
  author  = {Klett, Marcel and Hansmann, Philipp and Schäfer, Thomas},
  journal = {Frontiers in Physics},
  title   = {Magnetic properties and pseudogap formation in infinite-layer nickelates: Insights from the single-band {Hubbard} model},
  year    = {2022},
  volume  = {10},
  doi     = {10.3389/fphy.2022.834682}
}

@Article{Lihm2026,
  author  = {Lihm, Jae-Mo and Kiese, Dominik and Lee, Seung-Sup B. and Kugler, Fabian B.},
  journal = {Proceedings of the National Academy of Sciences},
  title   = {The finite-difference parquet method: Enhanced electron-paramagnon scattering opens a pseudogap},
  year    = {2026},
  number  = {10},
  pages   = {e2525308123},
  volume  = {123},
  doi     = {10.1073/pnas.2525308123}
}

@Article{Emery1995,
  author  = {Emery, V. J. and Kivelson, S. A.},
  journal = {Nature},
  title   = {Importance of phase fluctuations in superconductors with small superfluid density},
  year    = {1995},
  number  = {6521},
  pages   = {434--437},
  volume  = {374},
  doi     = {10.1038/374434a0}
}

@Article{Zhang2020,
  author  = {Zhang, Ya-Hui and Sachdev, Subir},
  journal = {Phys. Rev. Res.},
  title   = {From the pseudogap metal to the {Fermi} liquid using ancilla qubits},
  year    = {2020},
  number  = {2},
  pages   = {023172},
  volume  = {2},
  doi     = {10.1103/PhysRevResearch.2.023172}
}

@Article{Mascot2022,
  author  = {Mascot, Eric and Nikolaenko, Alexander and Tikhanovskaya, Maria and Zhang, Ya-Hui and Morr, Dirk K. and Sachdev, Subir},
  journal = {Phys. Rev. B},
  title   = {Electronic spectra with paramagnon fractionalization in the single-band {Hubbard} model},
  year    = {2022},
  number  = {7},
  pages   = {075146},
  volume  = {105},
  doi     = {10.1103/PhysRevB.105.075146}
}

@Article{Ayres2021,
author={Ayres, J.
and Berben, M.
and {\v{C}}ulo, M.
and Hsu, Y.-T.
and van Heumen, E.
and Huang, Y.
and Zaanen, J.
and Kondo, T.
and Takeuchi, T.
and Cooper, J. R.
and Putzke, C.
and Friedemann, S.
and Carrington, A.
and Hussey, N. E.},
title={Incoherent transport across the strange-metal regime of overdoped cuprates},
journal={Nature},
year={2021},
month={Jul},
day={01},
volume={595},
number={7869},
pages={661-666},
issn={1476-4687},
doi={10.1038/s41586-021-03622-z},
url={https://doi.org/10.1038/s41586-021-03622-z}
}

@article{Patel2024,
author = {Aavishkar A. Patel  and Peter Lunts  and Subir Sachdev },
title = {Localization of overdamped bosonic modes and transport in strange metals},
journal = {Proceedings of the National Academy of Sciences},
volume = {121},
number = {14},
pages = {e2402052121},
year = {2024},
doi = {10.1073/pnas.2402052121},
URL = {https://www.pnas.org/doi/abs/10.1073/pnas.2402052121}}

\clearpage

\thispagestyle{empty}

\setcounter{equation}{0}
\setcounter{figure}{0}
\setcounter{page}{1}

\renewcommand{\theequation}{S\arabic{equation}}
\renewcommand{\thefigure}{S\arabic{figure}}
\renewcommand{\thepage}{S\arabic{page}}

\setcounter{secnumdepth}{2} 
\renewcommand{\thefigure}{S\arabic{figure}}
\setcounter{figure}{0}
\setcounter{section}{0}
\setcounter{equation}{0}
\renewcommand{\thesection}{S-\Roman{section}}
\renewcommand{\theequation}{S\arabic{equation}}
%

%
\title{Supplemental Material for 
``{Quantum criticality in the two-dimensional Hubbard model}''}

\date{\today}
\maketitle

\section{Model and Method} \label{app:model&method}
The model we use to describe the cuprate physics is the Hubbard model \cite{Hubbard1963_model,Qin2022_HM}. Its lattice Hamiltonian reads
\begin{align}
    \nonumber H &=  \sum_{\br\br'\sigma}  -t_{\br\br'}\,c_{\br\sigma}^\dag c^{\pdag}_{\br'\sigma} -  \mu\sum_{\br\sigma} n_{\br\sigma} +  U\sum_{\br} n_{\br\uparrow} n_{\br\downarrow}\\
    &= \sum_{\bk\sigma} (\epsilon_{\bk}-\mu)\,n_{\bk\sigma}  +  U\sum_{\br} n_{\br\uparrow} n_{\br\downarrow}\ ,
\end{align}
where $c_{\br\sigma}^{(\dag)}$ are creation and annihilation operators in real space, and $n_{\br\sigma} = c_{\br\sigma}^\dag c^{\pdag}_{\br\sigma}$ denotes the number operator. Indices $\bk$ refer to  momentum space. The first, second and third terms represent the kinetic energy with hopping matrix elements $t_{\br\br'}$, the energy shift due to the chemical potential $\mu$, and the local on-site Coulomb interaction $U$, respectively. The kinetic energy contains only nearest-neighbor ($t$) and next-nearest-neighbor ($t'$) hopping terms, leading to the band dispersion $\epsilon_{\bk}=-2t(\cos k_x +\cos k_y) - 4t'\cos k_x \cos k_y$ and $\mu$ is used to control the doping $p$. Throughout, we use the parameters $t=1$, $U=7t$, focusing on $t'/t = -0.3$ in the main text and exploring $t'/t = \{0, -0.15, -0.2, -0.25, -0.3\}$ in \Fig{fig:t'_doping}. The temperature is set to $T=10^{-10}t$.

\begin{figure}[t!]
    \centering
    \includegraphics[width=1\linewidth]{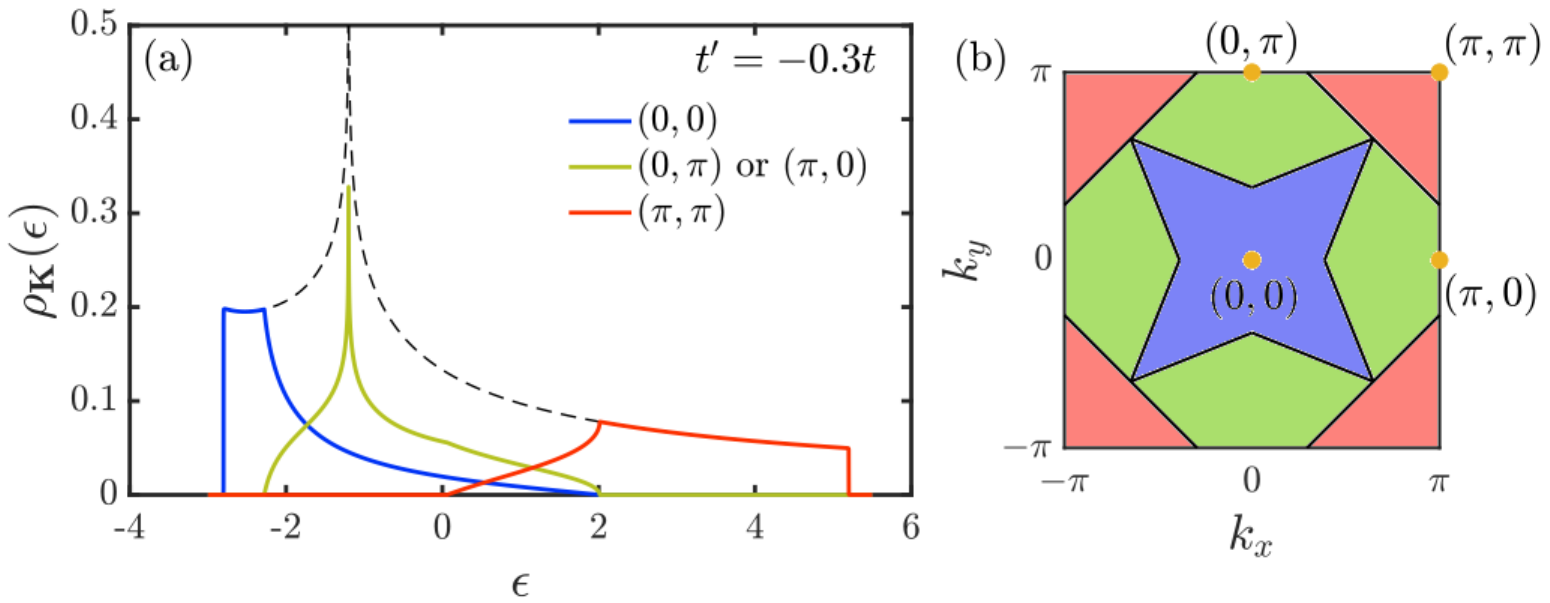}
    \caption{(a) Patch resolved density of states (DoS) $\rho_{\bK}(\epsilon)$ for $t'=-0.3t$. The dashed line indicates the sum of the four patches. (b) Corresponding patch geometry, referred to as ``star geometry''.}
    \label{fig:rhoK}
\end{figure}

We solve the model by means of DCA \cite{Hettler1998_DCA,Hettler2000_DCA,Maier2005_clusters,Gull2010_patching-pg,Potthoff2016_cDMFT}, due to mapping on a 2$\times$2 cluster impurity
\begin{align}
    \nonumber H_{\mr{imp}} &=  \sum_{\bR\bR'\sigma}  -\overline{t}_{\bR\bR'}\, c^\dagger_{\bR\sigma}c^{\pdag}_{\bR'\sigma} -  \mu \sum_{\bR\sigma} n_{\bR\sigma} +  U\sum_{\bR} n_{\bR\uparrow} n_{\bR\downarrow} \\
    &= \sum_{\bK\sigma} (\epsilon_{\bK}-\mu) n_{\bK\sigma} +  U\sum_{\bR} n_{\bR\uparrow} n_{\bR\downarrow} \ ,
\end{align}    
where $\bR$ denotes cluster impurity sites and $\bK$ the momentum averaged patches of the Brillouin zone (BZ) of the lattice model. On the impurity level the diagonal representation of the annihilation/creation operators $c^{(\dagger)}_{\bK\sigma} = \sum_{\bR} V^{\pdag}_{\bK\bR}c^{(\dagger)}_{\bR\sigma}$ are found by diagonalization $V_{\bK\bR} \overline{t}_{\bR\bR'} =\lambda_{\bK\bK'} V_{\bK'\bR'}$. The energies
\begin{gather}
    \epsilon_{\bK} = \int \mr{d}\epsilon \,\rho_{\bK}(\epsilon) \,\epsilon \ , \quad \rho_{\bK}(\epsilon) = \int_{V_{\bK}}\frac{\mr{d}\bk}{V_{\bK}}\delta(\epsilon-\epsilon_{\bk})\ , 
\end{gather}
where we integrate over momentum patches $V_{\bK}$ specified by the ``star patching'' geometry of \cite{Gull2010_patching-pg} and replotted in \Fig{fig:rhoK}(b). Note that nodal point $\bk=(\pi/2,\pi/2)$ is included in patch $\bK=(0,0)$. The corresponding patch density of states, $\rho_{\bK}(\epsilon)$, for $t'=-0.3t$ are shown in \Fig{fig:rhoK}(a). The dashed line indicates the sum of the four patches.

A NRG impurity solver \cite{Wilson1975_NRG,Krishna-murthy1980_NRG,Anders2005_NRG,Bulla2008_NRG,Lee2016_NRG}
is employed to evaluate the self-energy $\Sigma_{\bK}(z)$ by the means of the improved estimator approach \cite{Kugler2022_SEtrick}. Within NRG we interleave all four cluster momenta \cite{Mitchell2014_iNRG,Stadler2016_iNRG}, where the antinodal patches are symmetric. Therefore, references to patch $(0,\pi)$ also refer implicitly to $(\pi,0)$. We use the QSpace library \cite{Weichselbaum2012a_QSpace,Weichselbaum2012b_QSpace,Weichselbaum2020_QSpace,Weichselbaum2024_QSpace,Weichselbaum2024_QSpaceCode} to keep track of symmetries. An U(1) charge and SU(2) spin symmetry is enforced on the system prohibiting symmetry-breaking and the emergence of superconductivity. Within the computation we use a discretization $\Lambda=8$, no $z$-shifting, and kept multiplets of $N_{\mr{keep}}\leq 3\cdot10^4$. We double-checked the convergence in $N_{\mr{keep}}$ via $N_{\mr{keep}}\leq 10^5$ computations observing essentially no change in the spectral behavior of the correlators.

We use a density od states (DoS) integration for a stable computation of the local Green's function
\begin{align}
    \label{eq:G_LatInt}
    G_{\bK}(\z) = \int \frac{\mr{d}\epsilon \ \rho_{\bK}(\epsilon)}{\z +\mu - \epsilon - \Sigma_{\bK}(\z)} \, , 
\end{align}
where $z=\omega+\mr{i0^+}$. To update the hybridization $\Delta_{\bK}(\z)$ for the next DCA iteration we use an asymmetric estimator approach (\App{app:AsyEst_LatInt}), 
\begin{align}
    \label{eq:AsyEst}
    \Delta_{\bK}(\z)  = \frac{F_{\bK}(\z)}{G_{\bK}(\z)} -  \epsilon_{\bK}\,,\ \, 
    F_{\bK}(\z)  =  \int  \frac{\mr{d}\epsilon \ \rho_{\bK}(\epsilon)\ \epsilon}{\z  + \mu  -  \epsilon  -  \Sigma_{\bK}(\z)} \, . 
\end{align}

\subsection{Asymmetric estimator for lattice integration}
\label{app:AsyEst_LatInt}

In this section we discuss the computation of the hybridization, analogous to an asymmetric estimator approach within the DMFT context \cite{Kugler2022_SEtrick,Backes2022_genCavity}. The DCA approximation of the local lattice Green's function is
\begin{align}
    \label{eq:GKk}
    G_{\bK}(\bk,\z) = \frac{1}{\z +\mu -\epsilon({\bk}) - \Sigma_{\bK}(\z)} \, .
\end{align}
Rewriting and averaging over the patch volume $V_{\bK}$ gives
\begin{align}
    \label{eq:AsyEst_LatInt_intermediate}
    \big[\z +\mu - \Sigma_{\bK}(\z)\big]G_{\bK}(\z) - F_{\bK}(\z) = 1 \, ,
\end{align}
where
\begin{subequations}
\begin{align}
    G_{\bK}(\z)&=\int_{V_{\bK}}\frac{\mr{d}\bk}{V_{\bK}}G_{\bK}(\bk,\z) \, ,\\
    F_{\bK}(\z) &= \int_{V_{\bK}}\frac{\mr{d}\bk}{V_{\bK}}\epsilon({\bk})\,G_{\bK}(\bk,\z)\, .
\end{align}
\end{subequations}
Expressing these integrals as density-of-states (DoS) integrations yields Eqs.~\eqref{eq:G_LatInt} and \eqref{eq:AsyEst}.

Adding the identity $0=\epsilon_{\bK} + \Delta_{\bK}(\z)-\epsilon_{\bK} - \Delta_{\bK}(\z)$ to the first term of Eq.~\eqref{eq:AsyEst_LatInt_intermediate} gives
\begin{align}
    \big[G^{-1}_{\bK}(\z) + \epsilon_{\bK} + \Delta_{\bK}(\z)\big] G_{\bK}(\z) - F_{\bK}(\z) = 1 \, ,
\end{align}
where $G^{-1}_{\bK}(\z)=\z +\mu -\epsilon_{\bK} - \Delta_{\bK}(\z)-\Sigma_{\bK}(\z)$. Rearranging this expression leads to the asymmetric estimator for the hybridization, Eq.~\eqref{eq:AsyEst}.

Eq.~\eqref{eq:AsyEst} provides a more stable computation of the hybridization than the Dyson equation
\begin{align}
    \label{eq:Dyson_LatInt}
    \Delta_{\bK}(\z) = \z +\mu -\epsilon_{\bK} -\Sigma_{\bK}(\z) -G^{-1}_{\bK}(\z) \, .
\end{align}
Inverting numerical zeros of $G_{\bK}(\z)$ and subtracting these from $\Sigma_{\bK}(\z)$ in Eq.~\eqref{eq:Dyson_LatInt} leads to large peak heights, that scale with numerical errors. Consequently, this approach can produce large errors near the poles of $\Sigma_{\bK}(\z)$. In contrast, these errors cancel in the quotient form of Eq.~\eqref{eq:AsyEst}, keeping numerical errors small.

\section{Susceptibilities} \label{app:sus}
Here, we discuss the computation and behavior of various susceptibilities $\chi''[O](\omega)$ computed on the cluster level using our NRG impurity solver. Further, we provide an example of the energy scale extraction from $\chi''[O](\omega)$.

\subsection{Computed susceptibilities}
We have computed dynamical suceptibilities
$\chi[O](\omega)$, 
defined via \Eq{eq:chi}, for the following operators: 
\begin{subequations}
\begin{align}
    \nonumber C_{\bQ} &= c^{\dagger}_{\bK+\bQ,\sigma} \delta^{\pdag}_{\sigma\sigma'} c^{\pdag}_{\bK,\sigma'} & (\text{momentum charge}) \\
    \nonumber S^{z}_{\bQ} & = c^{\dagger}_{\bK+\bQ,\sigma} \sigma^{z}_{\sigma\sigma'} c^{\pdag}_{\bK,\sigma'} & (\text{momentum spin})
\end{align}
\end{subequations}
where $\bK$ and $\bQ$ are cluster momenta, and 
\begin{subequations}
\begin{align}
    \nonumber S^{\uparrow\pm\downarrow}_{d} &= (c^{\dagger}_{\delta\bD,\sigma}\pm c^{\dagger}_{\bZero,\sigma})\sigma^{z}_{\sigma\sigma'} (c^{\pdag}_{\delta\bD,\sigma'}\pm c^{\pdag}_{\bZero,\sigma'}) 
    \hspace{-1cm} & (\text{stag. spin})\\
    \nonumber P_{s} &= c^{\pdag}_{\bZero,\sigma} \mr{i}\sigma^y_{\sigma\sigma'} c^{\pdag}_{\bZero,\sigma'} & (s\text{-wave})\\
    \nonumber P_{x} &= c^{\pdag}_{\delta\bX,\sigma} \mr{i}\sigma^y_{\sigma\sigma'} c^{\pdag}_{\bZero,\sigma'} & (\text{NN singlet})\\
    \nonumber P_{d} &= c^{\pdag}_{\delta\bD,\sigma} \mr{i}\sigma^y_{\sigma\sigma'} c^{\pdag}_{\bZero,\sigma'} & (\text{NNN singlet})\\
    \nonumber P_{x^2\pm y^2} &= \big(c^{\pdag}_{\delta\bX,\sigma} 
    \pm c^{\pdag}_{\delta\bY,\sigma} \big)\mr{i}\sigma^y_{\sigma\sigma'} c^{\pdag}_{\bZero,\sigma'}& (\text{singlet})\\
    \nonumber I_{x} & = -\mr{i} t e( c^{\dagger}_{\delta\bX,\sigma} c^{\pdag}_{\bZero,\sigma}  - \mathrm{h.c.}) & (x\text{-current})\\
    \nonumber I_{d} & = -\mr{i} t' e( c^{\dagger}_{\delta\bD,\sigma} c^{\pdag}_{\bZero,\sigma}  - \mathrm{h.c.}) & (\text{diagonal current})
\end{align}
\end{subequations}
where $\delta\bX$, $\delta\bY$ and $\delta\bD= \delta\bX+\delta\bY$ denote real-space displacements relative to cluster site $\boldsymbol{0}$. Throughout, sums over repeated spin indices are implied, and $\sigma^z$ and $\sigma^y$ are Pauli matrices.
\begin{figure}[t!]
    \centering
    \includegraphics[width=1\linewidth]{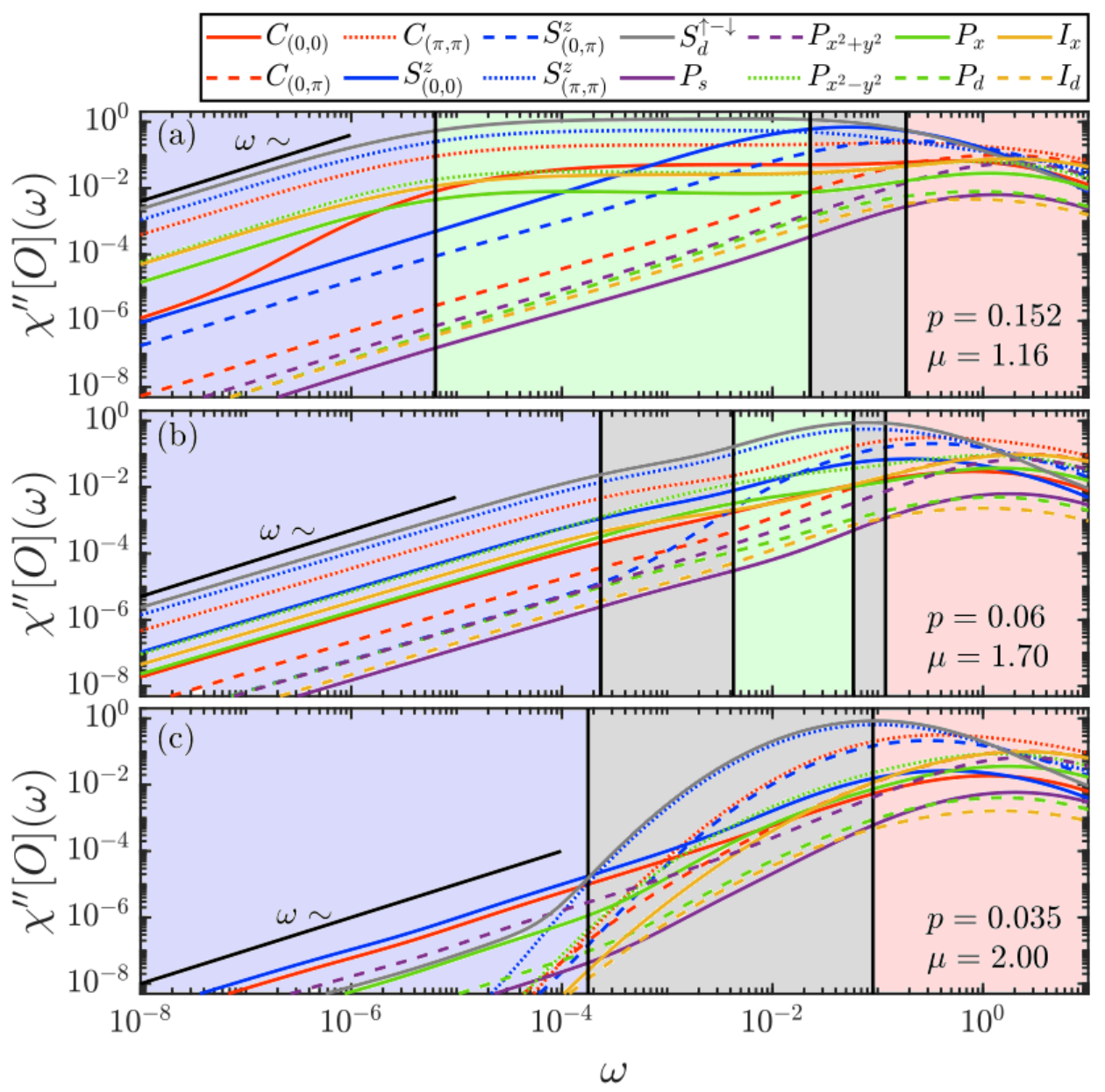}
    \caption{Spectral parts $\chi''[O](\omega)$ of the 
    susceptibilities of various operators $O$ with phase regimes corresponding to \Fig{fig:t'-0.3_phasedia} for (a) $p \gtrsim \pStar$, (b) $\pLS \lesssim p \lesssim \pStar$, and (c) $\pSig <  p < \pLS$.
    Color shadings correspond to the respective areas in \Fig{fig:t'-0.3_phasedia}(a).}
    \label{fig:chi_various}
\end{figure}
\begin{figure}[t!]
    \centering
    \includegraphics[width=1\linewidth]{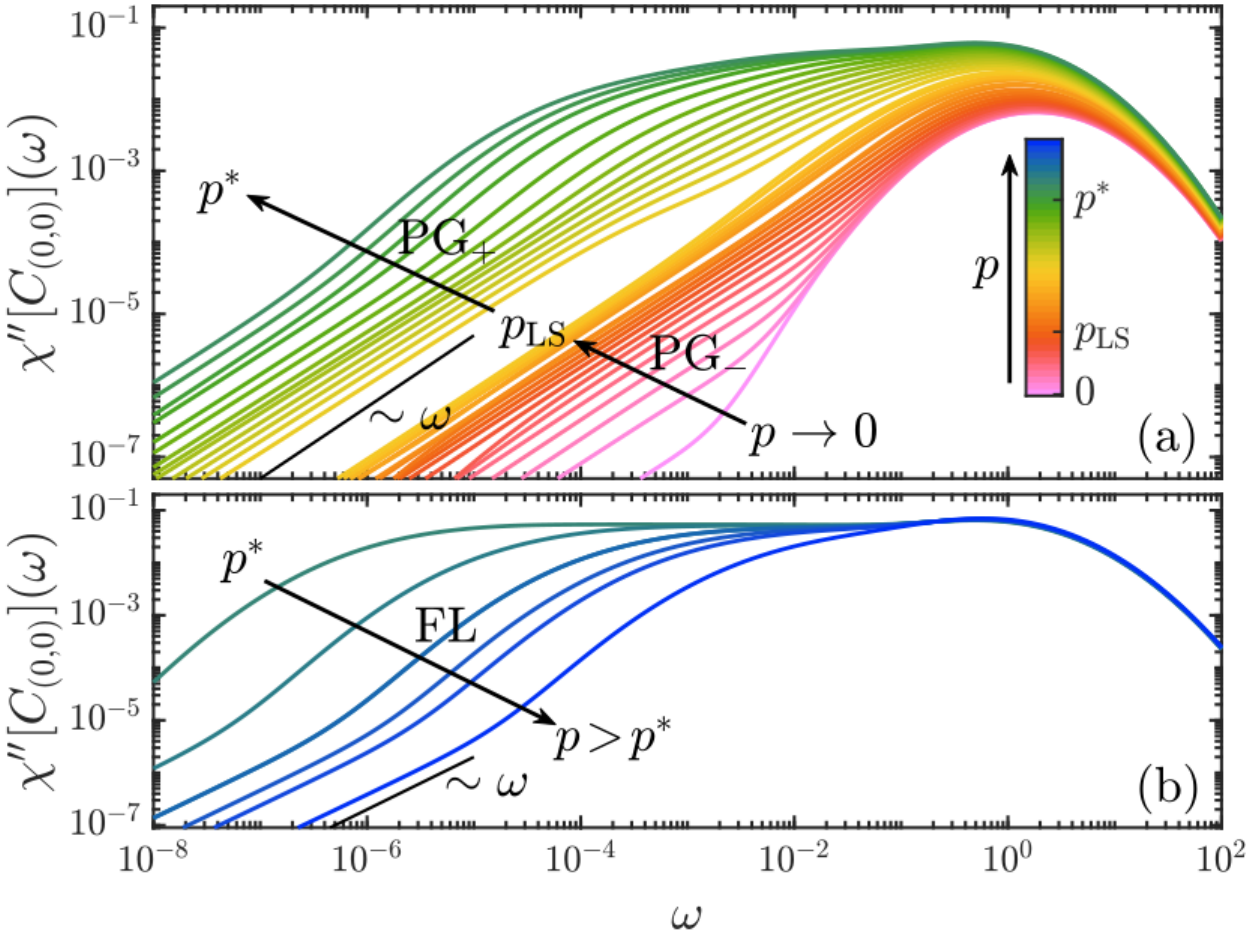}
    \caption{Spectral part of $\chi[C_{(0,0)}](\omega)$ shown on a log–log scale. Doping evolution from (a) $p \rightarrow 0$ (pink) to $\pStar$ (green), and (b) from $\pStar$ (green) to $p > \pStar$ (blue). Arrows indicate increasing $p$, tracing the progression through the corresponding phases.
    }
    \label{fig:chiC_doping}
\end{figure}

Figure \ref{fig:chi_various} presents the spectral parts $\chi''[O](\omega)$ of various such correlators for three values of doping.  Several of them, such as $\chi''[S^{\uparrow-\downarrow}_{d}]$, exhibit plateaus that widen for $p \to \pStar$, similar to those
seen in \Fig{fig:chiP_doping} of the main text.
Others, such as $\chi''[S^z_{(0,0)}]$,
show no plateau at all; for $p\to 0$ [\Fig{fig:chi_various}(c)] they show faster-than-power-law decay for $\omega\to0$, 
and with increasing $p$ evolve directly to linear-in-$\omega$ behavior all the way up to $\omega \simeq \TNFL$
[\Figs{fig:chi_various}(a,b)]. 
We have not found any obvious connection between operators whose spectra exhibit plateaus. A detailed understanding of why the spectra of so many seemingly unrelated operators exhibit this feature is left for future work. 

Figure~\ref{fig:chiC_doping} displays the evolution with doping of the  correlator $\chi''[C_{(0,0)}](\omega)$ in a manner similar to Fig.~\ref{fig:chiP_doping}.  The doping increases from $p \rightarrow 0$ to $p > \pStar$, as indicated by the color gradient from pink to blue and the accompanying arrows. The positions of $\pLS$ and $\pStar$ are marked accordingly. 

\subsection{Extraction of energy scales}
\begin{figure}[t!]
    \centering
    \includegraphics[width=1\linewidth]{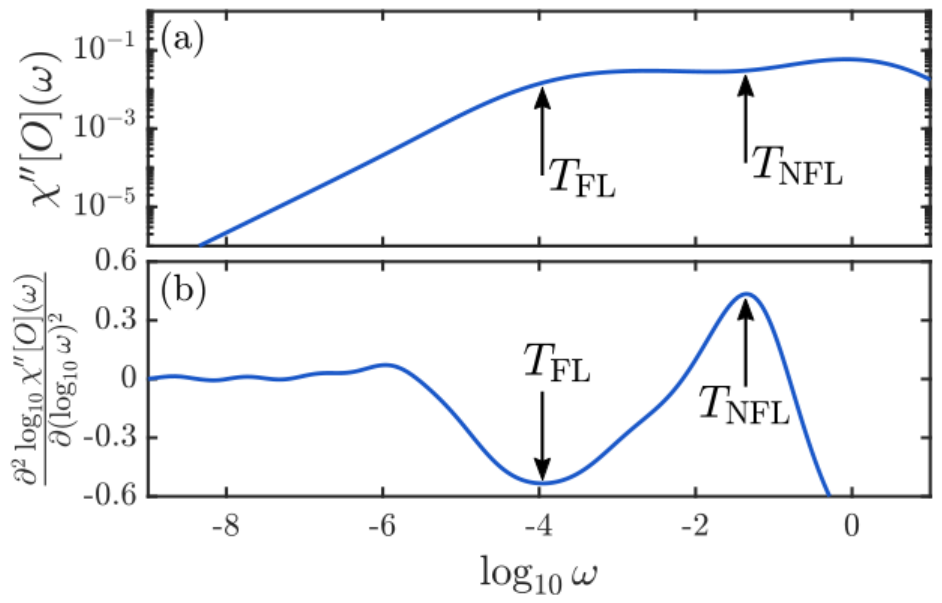}
    \caption{(a) A spectral function $\chi''[O](\omega)$ showing 
    a plateau. (b) Its second logarithmic derivative; we identify 
    $\TFL$ and $\TNFL$ with its extrema, which align with the plateau's ends.
    }
    \label{fig:chi_extraction}
\end{figure}

For spectral functions that exhibit a plateau,
we identify  the crossover scales $\TFL$ and $\TNFL$ 
with the left and right ``ends'' of the plateau. 
There, the function 
\begin{align}
    \frac{\partial^2\log_{10}\chi''[O](\omega)}{\partial(\log_{10}\omega)^2} 
\end{align}
takes on extremal values, signifying curvature changes in $\chi''[O](\omega)$ [Fig.~\ref{fig:chi_extraction}]. We identify the location of the first and second extremum with 
$\TFL$ and $\TNFL$, respectively, a robust and reliable manner for extracting these scales. 

\subsection{Static susceptibilities} \label{app:compressibility}

In \Fig{fig:t'-0.3_phasedia}(b) we present four static susceptibilities $\chi'(\omega = 0)$ that diverge for
$p$ approaching $\pStar$, implying a divergent \textit{local} response of the corresponding observable  at $\pStar$. 
This occurs for susceptibilities whose spectral parts $\chi''(\omega)$ exhibit a plateau whose extent 
increases towards $\omega \to 0$ for $p$ approaching $\pStar$. 

We use Kramers-Kronig to compute $\chi'(0)$ from the corresponding spectral function 
$\chi''(\omega)$ obtained from NRG, 
\begin{align}
    \label{eq:compress_app}
    \chi'[O] (\omega=0) = - \int\mr{d}\omega\,\frac{\chi''[O](\omega)}{\omega} \ .
\end{align}

In the charge sector, the static susceptibility is  directly related to the compressibility \cite{Peterson2010_compressibility},
\begin{align}
\kappa_T = \big(\partial n/\partial\mu\big)_T = - \lim_{{\bq} \to \boldsymbol{0}} \chi'[C_{\bq}](0) \, , 
\end{align}
where $C_\bq = \sum_{\bk,\sigma } c^{\dagger}_{\bk+\bq,\sigma} c^{\pdag}_{\bk,\sigma}$
involves operators defined on the lattice, not the cluster.
Viewing the cluster operator $C_{\bQ=(0,0)}$
as a proxy for $C_{\bq \to \boldsymbol{0}}$,
we may expect $\chi'[C_{\bQ = (0,0)}](0)$ to mimic the behavior 
of $\kappa_T$. In this regard, its discontinuous behavior
seen in  \Fig{fig:t'-0.3_phasedia}(b) is suggestive of a divergence
for $\kappa_T$. 
We expect that increased momentum resolution and closer proximity to $\pStar$ would reveal a clear divergence, defining a second-order phase transition at the quantum critical point, since $\kappa_T$ is defined as second derivative of the free energy.

\section{Patch Green's function, self-energy and hybridization} \label{app:SpecFunc}

In this appendix, we present and discuss the spectral  behavior of Green's function, self-energy and hybridization for individual patch momenta, and how the extracted critical dopings depend on $t'$.  

\subsection{Spectral behavior} \label{app:SpecFunc}
\begin{figure*}[t!]
    \centering
    \includegraphics[width=1\linewidth]{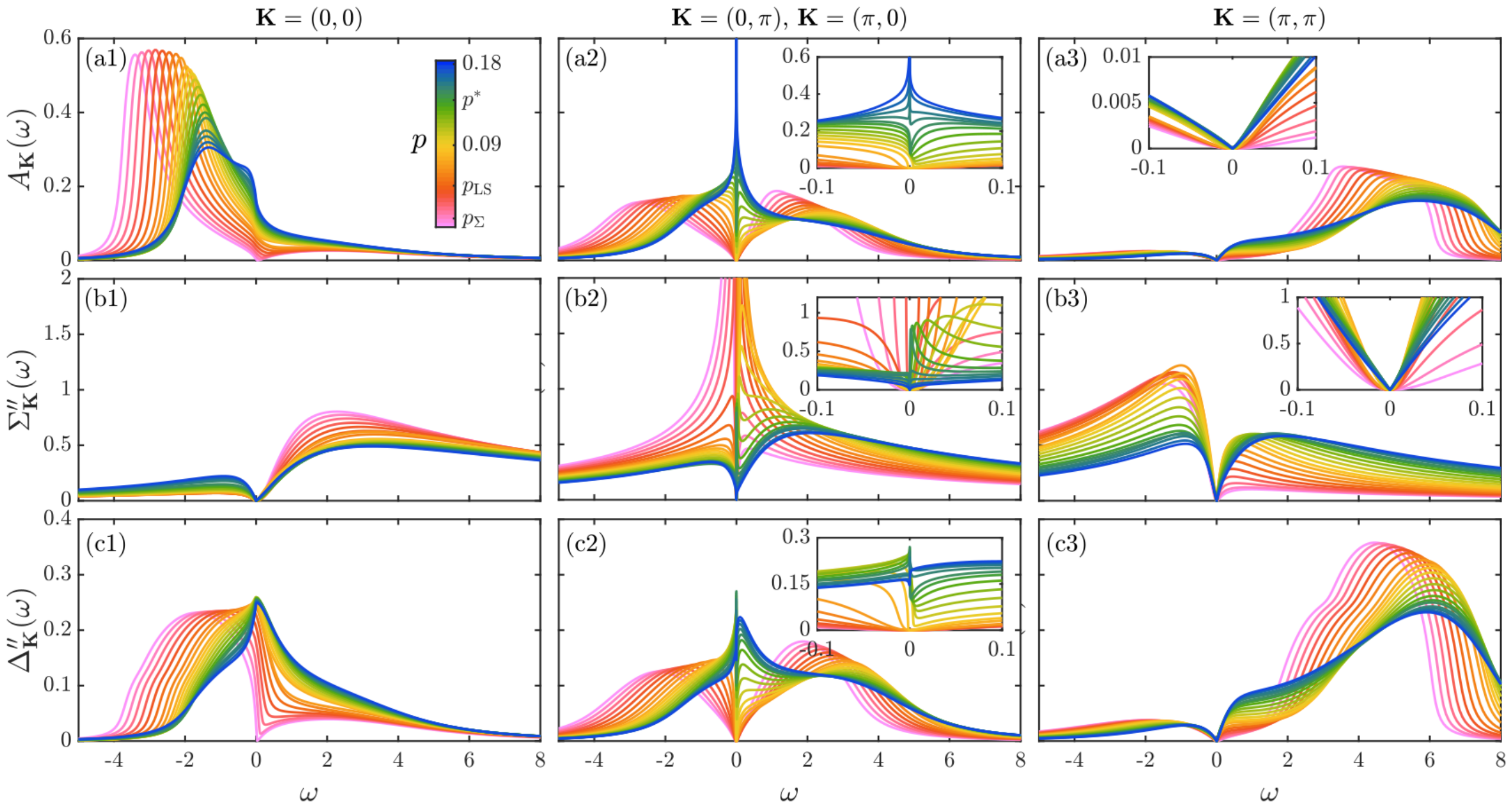}
    \caption{Spectral parts of (a) the Green's function, (b) the self-energy and (c) the hybridization, plotted vs.\ frequency for various dopings (color scale). Columns (1-3) present, respectively, patch momenta $\bK=(0,0)$, $\bK=(0,\pi)$ or $\bK=(\pi,0)$, and $\bK=(\pi,\pi)$. Insets zoom in to small frequencies around $\omega = 0$.
    }
    \label{fig:SpecFunc}
\end{figure*}
\begin{figure}[t!]
    \centering
    \includegraphics[width=1\linewidth]{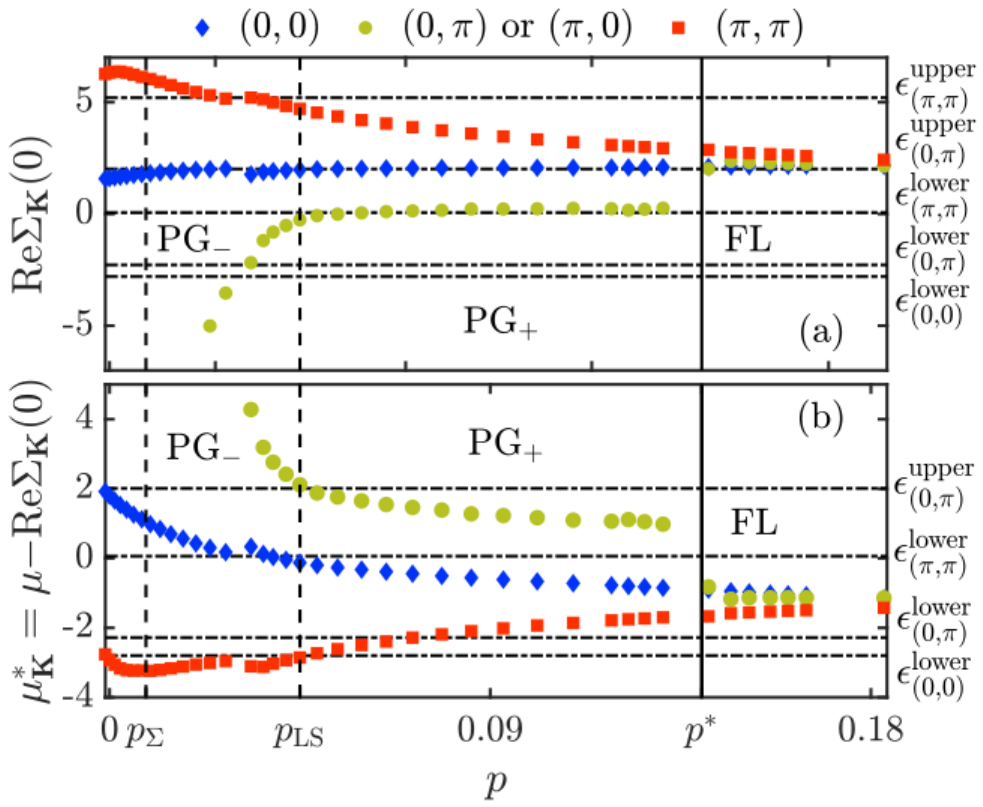}
    \caption{(a) Real part of the self-energy at $\omega=0$, $\mr{Re}\Sigma_{\bK}(0)$ as function of doping $p$. (b) $\mr{Re}\Sigma_{\bK}(0)$ shifted by the chemical potential $\mu$. Vertical black lines indicate the characteristic dopings 
    $\pSig$, $\pLS$ and $\pStar$; horizontal black lines mark $\epsilon^{\mr{lower}}_{\bK}$ and $\epsilon^{\mr{upper}}_{\bK}$, the lower and upper band edges of $\rho_{\bK}(\epsilon)$ in \Fig{fig:rhoK}, respectively. Here, $\epsilon^{\mr{upper}}_{(0,0)}=\epsilon^{\mr{upper}}_{(0,\pi)}$.
    }
    \label{fig:ReSE0}
\end{figure}

Here, we discuss the spectral behavior of the Green's function $G_\bK$, self-energy $\Sigma_\bK$ and hybridization $\Delta_\bK$ of individual patches. They all exhibit significant features at $\pStar$. 
Figure.~\ref{fig:SpecFunc} shows their spectral parts  $X_{\bK}''(\omega) = -\frac{1}{\pi}\mr{Im}\, X_{\bK}(\omega)$, corresponding to $A_\bK$, $\Sigma''_\bK$ and $\Delta''_\bK$, with color denoting doping $p$. As 
$p$ is decreased towards $\pStar$, the FL quasiparticle peak in $A_{(0,\pi)}$ (and $A_{(\pi,0)}$) weakens and then flips to a narrow gap-like feature
[\Fig{fig:SpecFunc}(a2)], while a self-energy peak 
in $\Sigma''_{(0,\pi)}$ emerges at $\omega = 0$ [\Fig{fig:SpecFunc}(b2)]. Simultaneously, the hybridization spectrum $\Delta''_{(0,\pi)}$ becomes strongly particle-hole asymmetric around $\omega = 0$
[\Fig{fig:SpecFunc}(c2)].

As $p$ decreases, these features intensify and shift slightly to positive frequencies until reaching $\pLS$. There, the small self-energy peak 
in $\Sigma''_{(0,\pi)}$ merges with the upper Hubbard band, forming a broad peak that evolves into the Mott divergence as $p \rightarrow 0$. Correspondingly, $A_{(0,\pi)}(\omega)$ and $\Delta''_{(0,\pi)}(\omega)$ develop a broad gap-like feature, which evolves with the self-energy peak into the full Mott gap. Finally, when 
$p$ drops below  $\pSig$, the location of 
the self-energy divergence changes sign from $\omega > 0$ to $\omega < 0$.

These features occur in the antinodal patches, which encompass the van Hove singularity of the non-interacting density of states [shown in \Fig{fig:rhoK}(a)]. The nodal patch spectral functions $A_{(0,0)}$ and $A_{(\pi,\pi)}$ correspond to the lower and upper Hubbard bands, respectively. The upper edge of the lower Hubbard band lies above $\omega = 0$, yielding finite spectral weight for $A_{(\pi,\pi)}(0)$ and contributing to metallic behavior. By contrast, $A_{(\pi,\pi)}$ remains gapped at all dopings, while its self-energy spectrum $\Sigma''_{(\pi,\pi)}$ shows no weight at $\omega = 0$, but exhibits strong particle-hole asymmetry in high-energy features. 
The gap for $A_{(\pi,\pi)}$ occurs because $\mu^{\ast}_{(\pi,\pi)} = \mu - \mr{Re} \Sigma_{(\pi,\pi)}(0)$, the effective chemical potential at $\omega = 0$, lies above the upper band edge $\epsilon^\mr{upper}_{(\pi,\pi)}$ of the patch density of states $\rho_{(\pi,\pi)}(\epsilon)$ [c.f. Fig.~\ref{fig:rhoK}(a)], 
so that  $\rho_{(\pi,\pi)}(\mu^{\ast}_{(\pi,\pi)}) = 0$, see Fig.~\ref{fig:ReSE0}(a).

The Mott transition occurs at the critical doping $\pStar$, where the self-energy pole in the antinodal patches emerges.  The full gap opening in the antinodal patches at $\pLS$ occurs because the effective chemical potential at $\omega = 0$, $\mu^{\ast}_{\bK} = \mu - \mr{Re} \Sigma_{\bK}(0)$, continuously shifts outside of the patch bandwith of the $\bK = (0,\pi)$ and $\bK = (\pi,0)$ patches,
so that $A_{\bK}(0)= \rho_{\bK}(\mu^{\ast}_{\bK}) = 0$,resulting in a full gap [see \Figs{fig:ReSE0}(a,b)]. The same occurs in the nodal patch as $p \to 0$, leading to a fully gapped Mott insulator. Neither $\pLS$ nor $p = 0$ is associated with a critical point.

The antinodal self-energy exhibits a discontinuity at $\pStar$, associated with the emergence of the pseudogap, while the nodal patches evolve smoothly with doping. The doping $\pLS$ marks the point where 
$\mu^{\ast}_{(0,\pi)}$ crosses the band edge of $\rho_{(0,\pi)}(\epsilon)$, opening a full gap in the antinodal patch.
This is consistent with the known patching dependence of the Hubbard model \cite{Gull2010_patching-pg}.

Notably, $\mu$ never shifts $\mr{Re}\Sigma_{\bK}(0)$ into the band edge regime of $\epsilon^{\mr{lower}}_{(\pi,\pi)}$. This explains why the $\bK=(\pi,\pi)$ patch remains gapped at all dopings, despite its relatively FL-like self energy. Interestingly, $\mu$ decreases with increasing doping $p$. While both the nodal $\bK=(0,0)$ and antinodal patches follow this trend, the behavior of the $\bK=(\pi,\pi)$ patch is opposite. This distinction is crucial: it prevents the Fermi surface from entering the $\bK=(\pi,\pi)$ patch at low doping and thereby enables the Mott insulating state at half-filling.

\subsection{Electronic quasiparticle} \label{app:QP}

\begin{figure}[t!]
    \centering
    \includegraphics[width=1\linewidth]{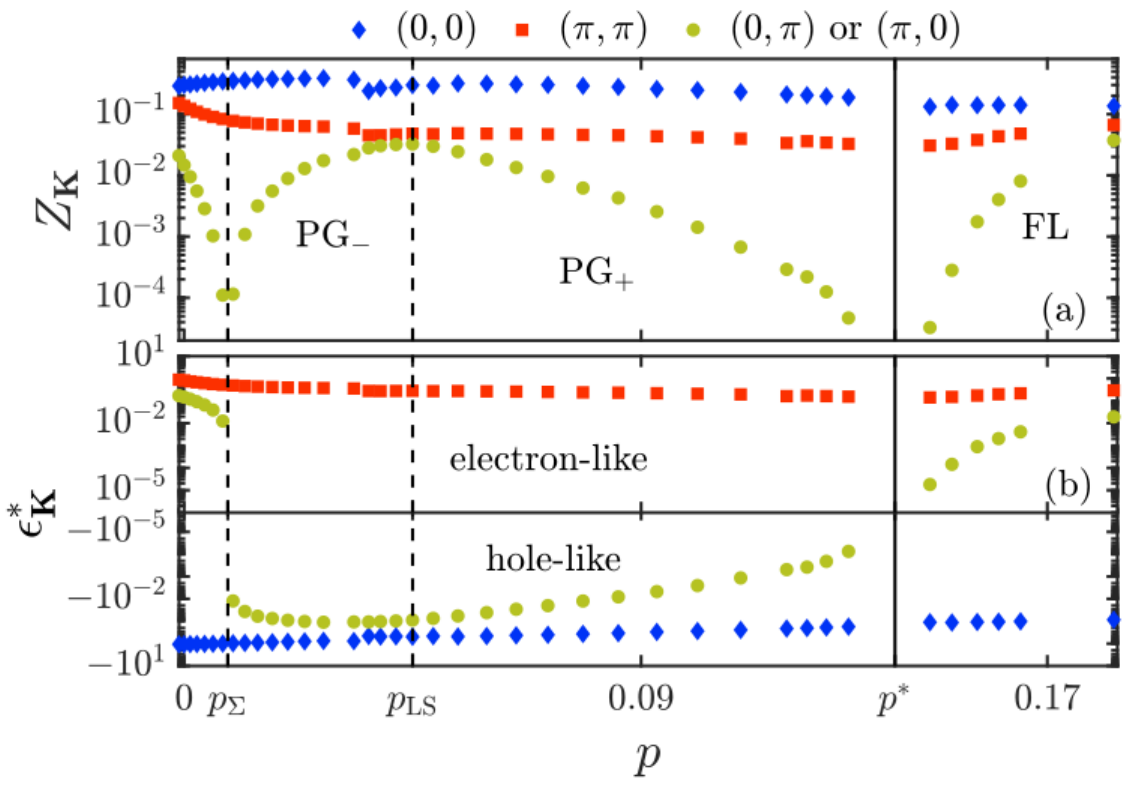}
    \caption{(a) Quasiparticle weight $Z_{\bK}$, and (b) energies $\epsilon^{\ast}_{\bK}$, as function of $p$ for momenta (blue) $\bK = (0,0)$, (green) $\bK = (0,\pi)$ or $\bK = (\pi,0)$, and (red) $\bK = (\pi,\pi)$.}
    \label{fig:ZK_EKstar_ReSE}
\end{figure}

In this section we focus on the electronic quasiparticle behavior as function of doping. Figure \ref{fig:ZK_EKstar_ReSE} shows the $p$-dependence of patch quasiparticle (QP) weights and energy, defined as
\begin{subequations}
\begin{align}
    Z_{\bK} &= \big( 1-\partial_{\omega}\Sigma_{\bK}'(\omega)\big|_{\omega = 0}\big)^{-1}\, ,\\
    \epsilon_{\bK}^{\ast} &= Z_{\bK}\, [\epsilon_{\bK}-\mu+\Sigma_{\bK}'(0)] \, , 
\end{align}
\label{eq:QP}
\end{subequations}
Colors blue, green, and red denote individual momenta.

The nodal quasiparticle weight $Z_{\bK}$ in \Fig{fig:ZK_EKstar_ReSE}(a) exhibit an approximately linear dependence on doping, compared to the pronounced non-monotonic behavior of their antinodal counterparts. On a logarithmic scale, $Z_{(0,\pi)}$ (and equivalently $Z_{(\pi,0)}$ vanishes at $\pStar$ and $\pSig$ due to the emergence of a self-energy pole at $\omega = 0$. Between these dopings, $Z_{(0,\pi)}$ displays a maximum at $\pLS$. 

Figure~\ref{fig:ZK_EKstar_ReSE}(b) shows the energies ${\epsilon}_{\bK}^{\ast}$, which yield information about the electronic QP character. Across the entire doping range, the $\bK = (0,0)$ and $\bK = (\pi,\pi)$ patches retain hole- and electron-like character, respectively, consistent with conventional Fermi liquid behavior. In contrast, the antinodal energies change sign at the critical dopings 
$\pStar$ and $\pLS$. For $\Sigma_{\bK}$, these sign changes are discontinuous, reflecting poles in the real part of the self-energy. By contrast, ${\epsilon}_{\bK}^{\ast}$ evolves continuously, albeit with a very steep crossover near $\pSig$. The shift from electron- to hole-like character is consistent with a transfer of coherence from the conventional electron-like Fermi liquid at $p>\pStar$ to the emergent composite Fermi liquid at $p<\pStar$.

\subsection{Dependence on next nearest neighbor hopping $t'$} \label{app:tprime}

In this section, we discuss the dependence of the phase diagram on the next nearest neighbor hopping $t'$. In the main text, we focus on the representative case $t'=-0.3t$, which captures all qualitative features of the phase diagram. Here, we briefly comment on how these features evolve as a function of $t'$.

\begin{figure}[t!]
    \centering
    \includegraphics[width=1\linewidth]{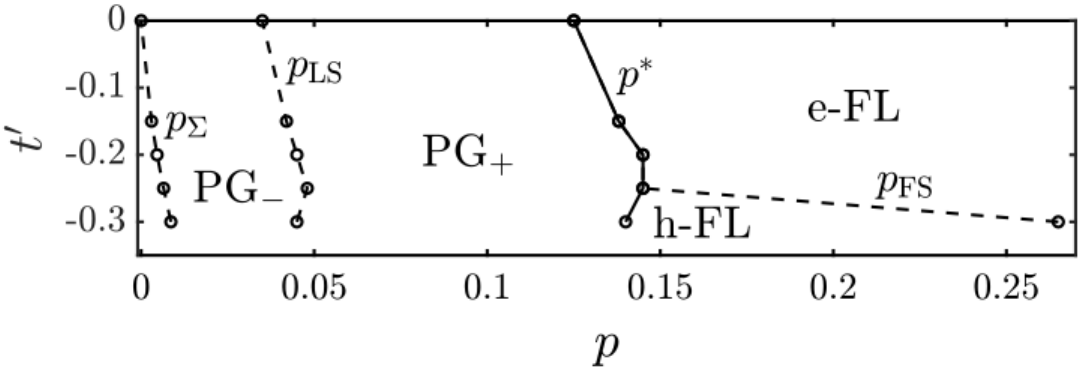}
    \caption{Dependence of various critical doping values on next-nearest-neighbor hopping $t'$ and doping $p$. The critical doping $\pStar$ marks the transition from the Fermi liquid (FL) to the pseudogap metal (PG).  Dashed lines $\pFS$ and $\pLS$ denote topology changes of the Fermi surface (FS) and Luttinger surface (LS), respectively. The line $\pSig$ indicates the doping below which the self-energy pole shifts from $\omega>0$ to $\omega < 0$.}
    \label{fig:t'_doping}
\end{figure}

In Fig.~\ref{fig:t'_doping}, we show the critical dopings as a function of $t'$. The line labeled $\pStar$ marks the transition between the Fermi liquid (FL) regime and the pseudogap (PG) metal. This transition is characterized by a reconstruction of the Fermi surface and the simultaneous vanishing of the Fermi liquid coherence scale in multiple response functions. The doping $\pSig$ denotes the point at which the self-energy pole shifts from $\omega>0$ for $p>\pSig$ to $\omega<0$ for $p<\pSig$. The dopings $\pFS$ and $\pLS$ indicate changes in the topology of the Fermi surface (FS) and the Luttinger surface (LS), respectively.

The critical dopings shown in Fig.~\ref{fig:t'_doping} are determined by characteristic changes in spectral behavior, which are equal to the $t'=-0.3t$ case presented in \Fig{fig:SpecFunc}.

For $t'\geq 0.25t$, the critical dopings $\pStar$ and $\pFS$ merge into a single transition. Apart from this merging, the qualitative behavior of the system remains unchanged across different values of $t'<0$. Therefore, the choice $t'=-0.3t$ provides a representative description, and the conclusions drawn in the main text apply generally to other negative values of $t'$.

\section{Momentum dependence of the Green's function}

In this section, we examine the doping evolution of the real part of the cluster Green’s function at $\omega=0$ across momentum and present the momentum dependence of the Green’s function and self-energy near the critical transition complementary to the main text.

\subsection{Fermi and Luttinger surface versus doping} \label{app:FS_noperiod}

\begin{figure*}[t!]
    \centering
    \includegraphics[width=1\linewidth]{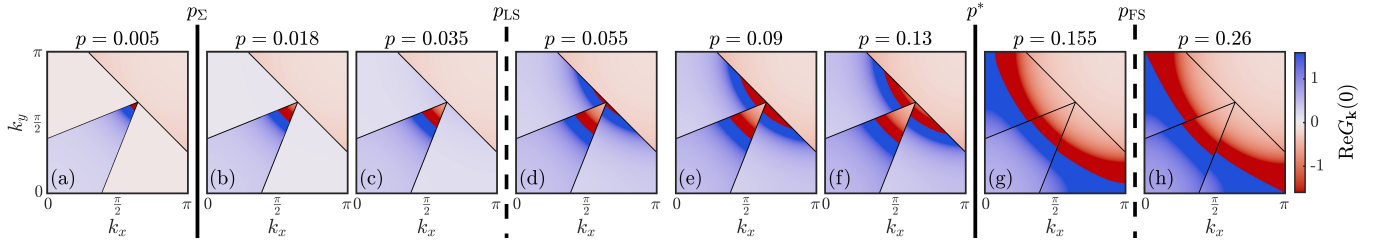}
    \caption{Real part of the Green's function at $\omega = 0$, $\mr{Re}\, G_{\bk}(0)$, 
   computed without $\mc{L}$-interpolation,  shown in a quadrant of the Brillouin zone for several values of doping $p$, increasing from (a) to (h). Black separators between panels mark characteristic dopings.
    Thin, straight black lines in each panel show  patch boundaries. The Fermi surface is associated with a sign change via a divergence (strong-blue to strong-red), a Luttinger surface with a sign change via a small discontinuity (light-blue to light-red) at the patch boundaries.  }
    \label{fig:FS_doping_app}
\end{figure*}

This section provides insight into the evolution of the Fermi and Luttinger surfaces as a function of $p$. To this end, we use the cluster results for the real part of the Green's function at $\omega=0$, $\mr{Re}\, G_{\bk}(0)$. 

The DCA result for the Green's function gives 
\begin{align}
    G_{\bk}(\omega) = \frac{1}{\omega+\mu-\epsilon_{\bk}- \Sigma_{\bk}(\omega)}\, ,
\end{align}
with  the self-energy given by $\Sigma_{\bk}(\omega)=\theta_{\bk,\bK}\Sigma_{\bK}(\omega)$, where $\theta_{\bk,\bK}$ is a patch-selective box function. 
Liouvillian interpolation replaces the latter  piecewise constant form by a smoothly momentum-dependent version,  obtained via a continued-fraction expansion \cite{Pelz2026_Linterp}. 

Figure~\ref{fig:FS_doping_app} shows the non-interpolated version of $\mr{Re} \, G_{\bk}(0)$.  A sign change via a divergence within a patch (strong blue to strong red) define the Fermi surface, while small, step-like sign changes across patch boundaries (light-blue to light-red) indicate a Luttinger surface. 

A Fermi surface topology change occurs at $\pFS\simeq0.25$, where the Fermi surface changes smoothly from electron-like to hole-like. 
As $p$ drops below $\pStar$ and decreases towards $\pLS$,
we observe an increasing mismatch between the Fermi surface of the nodal $\bK = (0,0)$ patch and the antinodal patches. This is accompanied by the emergence of a Luttinger surface at the antinodal patch edges toward $\bK = (\pi,\pi)$, driven by the PG$_+$ self-energy peak. In this regime, the Luttinger surface expands while the Fermi surface shrinks with decreasing doping.

When $p$ drops below $\pLS$, the Fermi surface in the antinodal patches disappears and a closed Luttinger surface forms as the PG$_+$ self-energy peak merges with high-energy features, producing a Mott divergence in the antinodal patch. This constitutes a reconstruction of 
the Luttinger surface topology from open to closed. However, this is a crossover, not a thermodynamic transition, since previous studies showed that its details depend on the patching scheme chosen \cite{Gull2010_patching-pg}. For example,
improving the $\bk$-resolution by using a larger cluster, may well lead to different behavior for the Luttinger surface.

When $p$ drops below $\pSig$, the real part of the Green’s function changes sign (positive to negative) in the antinodal patches, reflecting a sign change (positive to negative) in the location of the antinodal self-energy pole. However, Fermi surface and metallic behavior  survive in the $\bK = (0,0)$ patch throughout 
as $p$ drops below $\pLS$ and then $\pSig$, 
vanishing only as $p \to 0$. 

\begin{figure}[t!]
    \centering
    \includegraphics[width=1\linewidth]{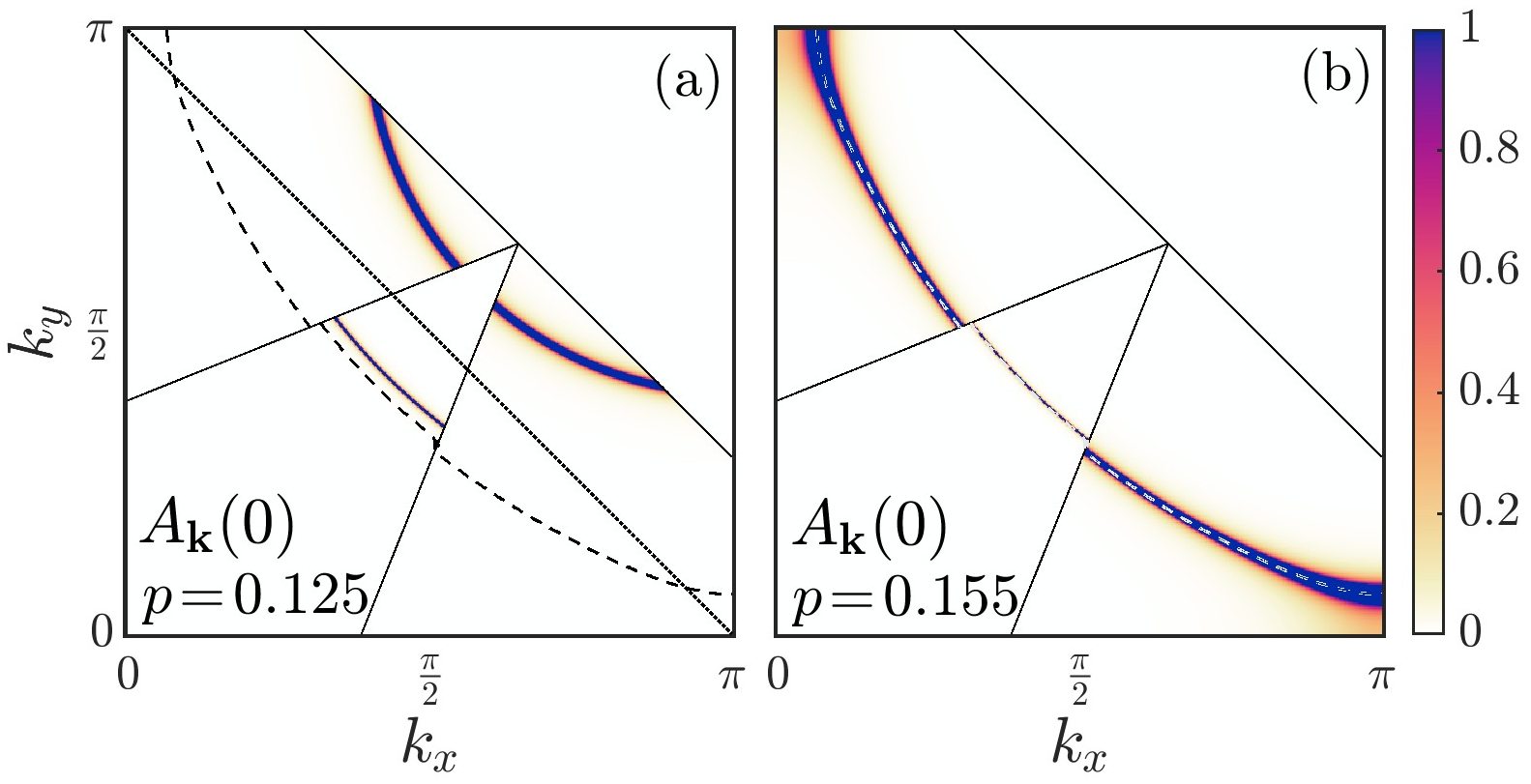}
    \caption{Momentum-resolved spectral functions 
    at $\omega=0$, $A_{\bk}(0)$, computed without $\mc{L}$-interpolation for (a) $p=0.125$ and (b) $p=0.155$. Thin solid black lines indicate patch boundaries. In (a), the dotted line  marks the antiferromagnetic zone boundary. In (b), the curved dashed line marks the locus of maximal $A_{\bk}(0)$; it is shown in (a), too, as a reference for the unreconstructed Fermi surface. Figures~\ref{fig:ARPES_A0}(c2,c1) show the $\mc{L}$-interpolated versions of (a,b).}
    \label{fig:A0_cluster}
\end{figure}

In the main text, \Figs{fig:ARPES_A0}(c1,c2) showed the
zero-frequency spectral function 
$A_{\bk}(0)$  for two doping values just 
above and below $\pStar$, computed using a momentum-smooth self-energy $\Sigma''_\bk$ obtained via Liouvillian interpolation \cite{Pelz2026_Linterp}. To complement that discussion, \Figs{fig:A0_cluster}(a,b)  shows $A_{\bk}(0)$ for the same two dopings,
now computed without Liouvillian interpolation, 
i.e.\ using  momentum-discontinuous patch self-energies
$\Sigma_\bK$. 
Whereas the  nodal patches 
in for $p> \pStar$ and $p< \pStar$ hardly differ, implying a smooth evolution across $\pStar$, the antinodal patches differ strongly, 
signaling a Fermi surface reconstruction across
$\pStar$. Note that for $p< \pStar$, the region of non-zero spectral weight in the antinodal patches no longer crosses the antiferromagnetic zone boundary (for $p> \pStar$, it did). The resulting Fermi surface 
represents a Fermi arc (albeit a discontinuous one,
reflecting DCA patching) in that the antinodal surface does not extend to the boundary of the Brillouin zone.
Figure~\ref{fig:ARPES_A0}(c1) shows the $\mc{L}$-interpolated (continuous) version of this Fermi arc.

\FloatBarrier
\clearpage

\end{document}